\documentclass[conference]{IEEEtran}
\IEEEoverridecommandlockouts
\usepackage{cite}
\usepackage{amsmath,amssymb,amsfonts}
\usepackage{algorithmic}
\usepackage{graphicx}
\usepackage{textcomp}
\usepackage{xcolor}
\usepackage{booktabs, multirow,array}
\usepackage{float}
\usepackage{subfig}

\def\BibTeX{{\rm B\kern-.05em{\sc i\kern-.025em b}\kern-.08em
    T\kern-.1667em\lower.7ex\hbox{E}\kern-.125emX}}
\begin{document}

\title{Joint Scatterer Localization and Material Identification Using Radio Access Technology
%hould not be used}
}
\makeatletter
\newcommand{\linebreakand}{%
  \end{@IEEEauthorhalign}
  \hfill\mbox{}\par
  \mbox{}\hfill\begin{@IEEEauthorhalign}
}
\makeatother
\author{\IEEEauthorblockN{Yi Geng}
	\IEEEauthorblockA{\textit{Ericsson R\&D} \\
		Nanjing, China \\
		yi.a.geng@ericsson.com}
	\and
	\IEEEauthorblockN{Deep Shrestha}
	\IEEEauthorblockA{\textit{Ericsson Research} \\
		Link{ö}ping, Sweden \\
		deep.shrestha@ericsson.com}
	\and
	\IEEEauthorblockN{Vijaya Yajnanarayana}
	\IEEEauthorblockA{\textit{Ericsson Research } \\
		Banglore, India \\
		vijaya.yajnanarayana@ericsson.com}
	\linebreakand
	\IEEEauthorblockN{Erik Dahlman}
\IEEEauthorblockA{\textit{Ericsson Research} \\
	Kista, Sweden \\
	erik.dahlman@ericsson.com}
	\and
		\IEEEauthorblockN{Ali Behravan}
	\IEEEauthorblockA{\textit{Ericsson Research} \\
		Kista, Sweden \\
		ali.behravan@ericsson.com}
}
\maketitle

\begin{abstract}
Cellular network technologies and radar sensing technologies have been developing in parallel for decades. Instead of developing two individual technologies, the 6G cellular network is expected to naturally support both communication and radar functionalities with shared hardware and carrier frequencies. In this regard, radio access technology~(RAT)-based scatterer localization system is one of the important aspects of joint communication and sensing system~(JCAS) that uses communication signals between transceivers to determine the location of scatterers in and around the propagation paths. In this article, we first identify the challenges of RAT-based scatterer localization system, then present single- and multiple-bounce reflection loss (RL) simulation results for three common building materials in indoor environments. We also propose two novel methods to jointly localize and identify the type of the scatterers in a rich scattering environment.
\end{abstract}


\section{Introduction}
Depending on whether the objects of interest have communication capability or not, the localization technologies are classified into active localization and passive localization \cite{b1}. Active localization systems use transmitted and received signals both from localization system and objects with communication capability to localize the objects. Almost all RAT-based localization systems are active localization (e.g., localizing user equipment~(UE) in a cellular network). Passive localization systems localize scatterers without communication capability by exploiting reflected signals induced by the scatterers (e.g., detecting a plane by using a bistatic radar).

Unlike the existing passive localization use cases, some emerging 6G localization applications such as robotic perception, virtual reality~(VR), digital twins, and three-dimensional (3D) digital mapping require scatterer information of not only the position but also the material \cite{b2}. By combining the scatterers' location information with their material information of an environment, a 3D digital map with another layer of material information can be generated. This supplementary information in the 3D digital map can find applications in VR games and emerging area of virtual tourism due to COVID-19-caused travel disruption. Current VR suits can give visual and audio illusions by simulating human senses of sight and hearing. However, with material information in 3D digital map, VR applications can generate additional human sensations like touch and smell. For example, a VR game player can feel the hardness, the temperature, and the odour of the virtual objects in a simulated environment through sensory feedback. Material sensing can be a key requirement for autonomous driving~(AD) also. Today the onboard radar systems of vehicles are capable of estimating precise position of obstacles around the vehicles. However, having an added information about the type of the material can aid in several ways. For example, by sensing the snow on the road, vehicle can tune its onboard electronics such as traction-control system or change to another safer route.

This article is organized as follows. In Section~II, we present the challenges of RAT-based scatterer localization system. In Section~III, we analyze RL of single- and double-bounce reflection for three common building materials in indoor environments. In Section~IV, we propose two novel methods all based on RL to trace radio trajectories, localize scatterers’ position, and identify scatterers' material in rich scattering environments. Also, in this section the performance of proposed methods is evaluated. To the best of the authors’ knowledge, this is the first work that shows how to localize scatterers and identify scatterers' material simultaneously by using wireless communication signals.

\begin{figure*}[t]
	\centering
	\subfloat[Single-bounce]{\label{fig:a}\includegraphics[width=0.65\columnwidth]{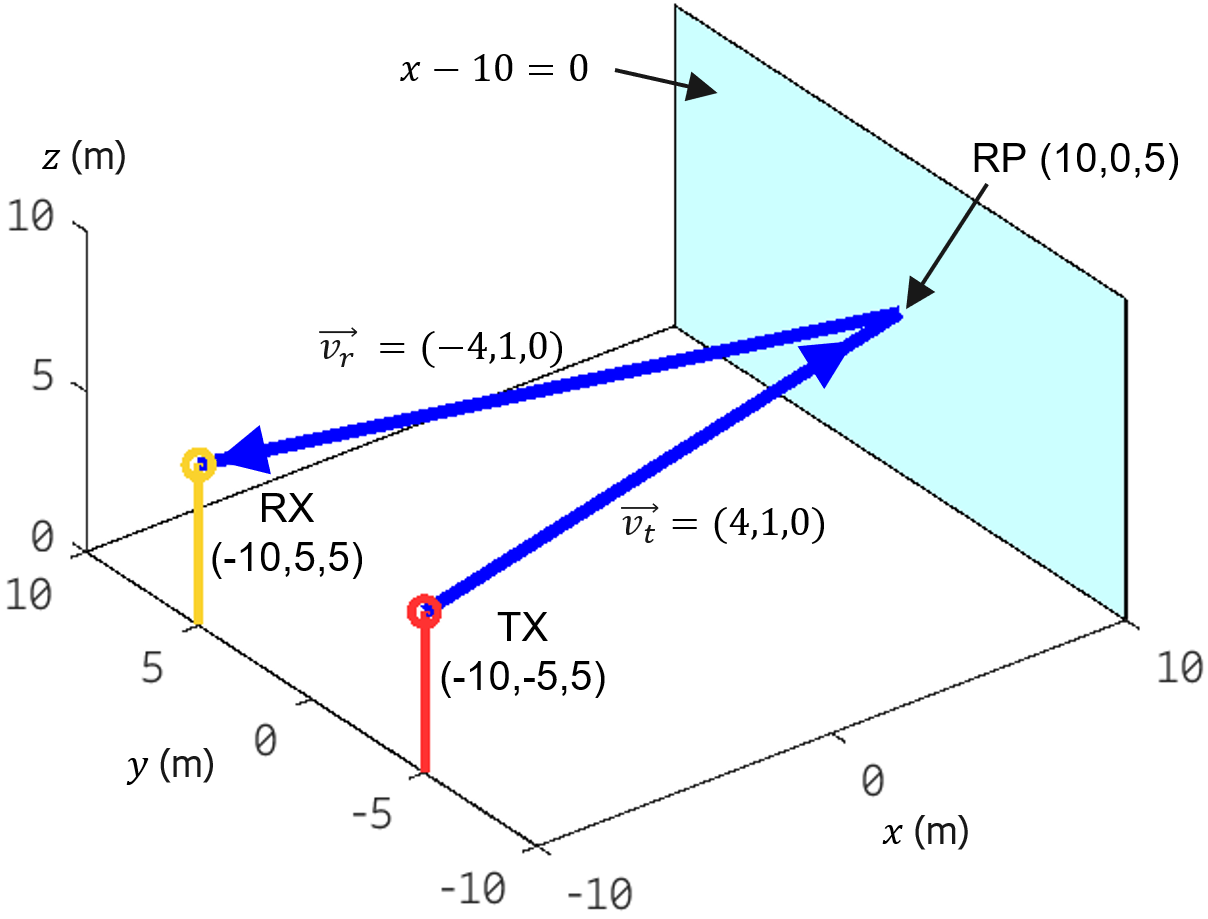}}\quad
	\subfloat[Double-bounce]{\label{fig:b}\includegraphics[width=0.65\columnwidth]{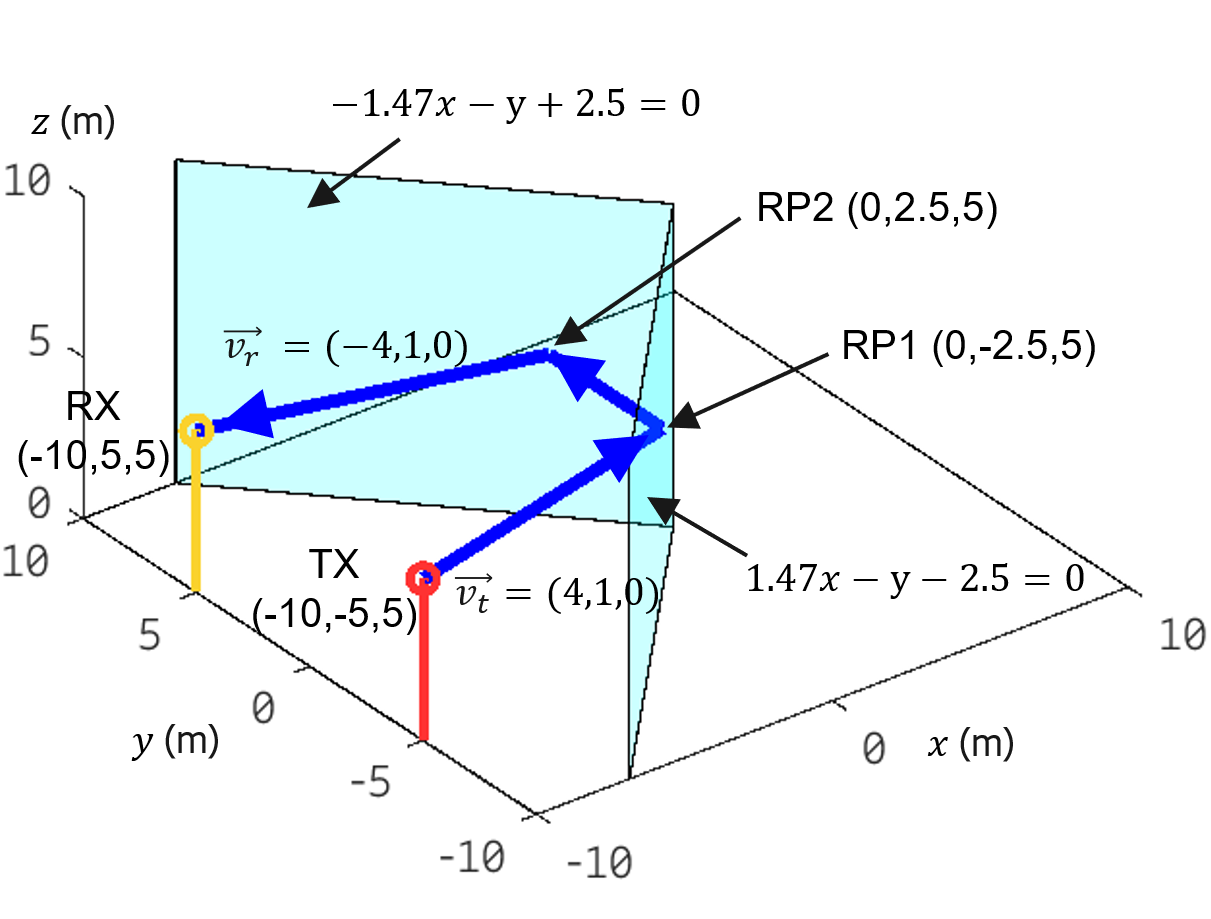}}\quad
	\subfloat[Triple-bounce]{\label{fig:c}\includegraphics[width=0.65\columnwidth]{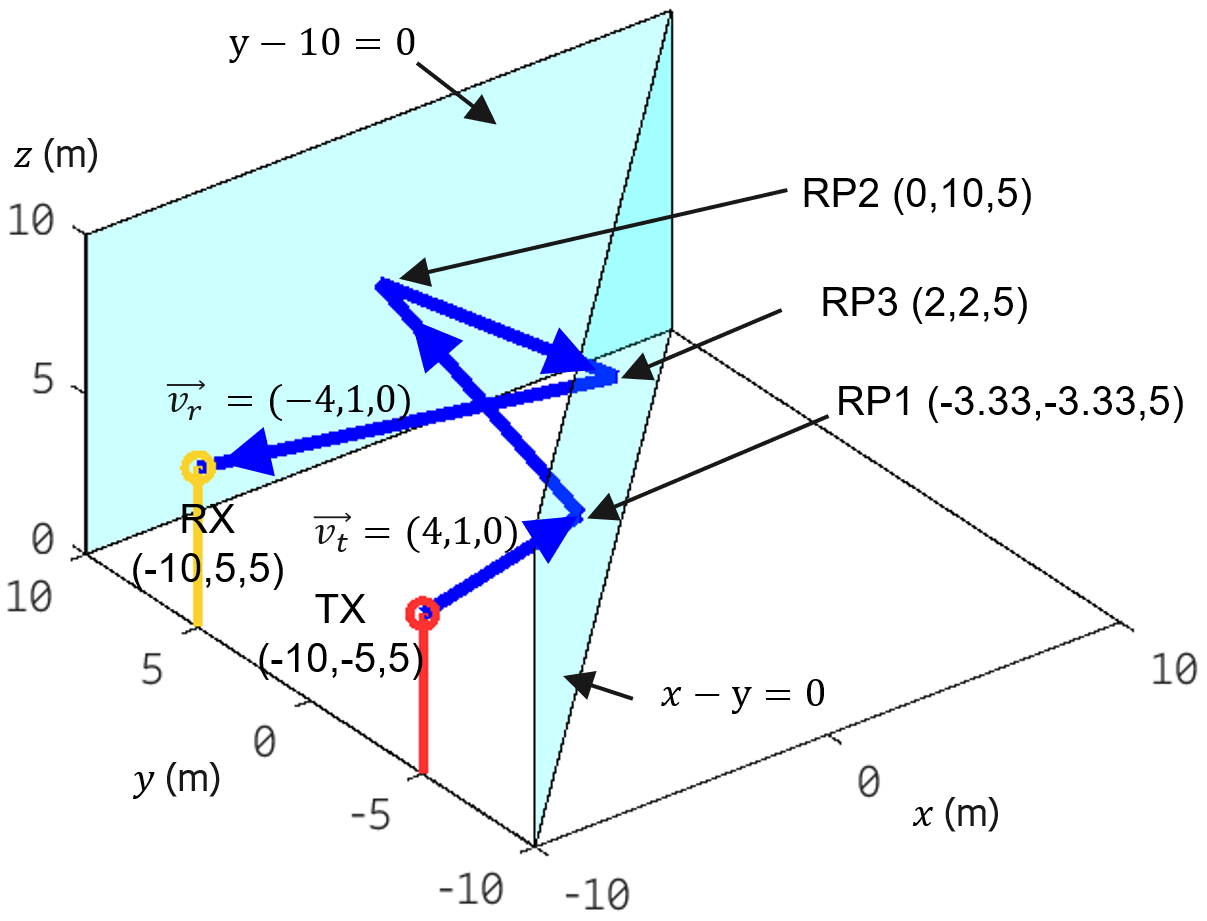}}\\	
	\caption{The possible trajectories of a radio ray transmitted from \mbox{TX(-10, -5, 5)} in the direction of vector (4, 1, 0) and received at \mbox{RX(-10, 5, 5)} in the direction of vector (-4, 1, 0).}
\end{figure*}

\begin{figure*}[t]
	\centering
	\subfloat[Trajectory A]{\label{fig:d}\includegraphics[width=0.8\columnwidth]{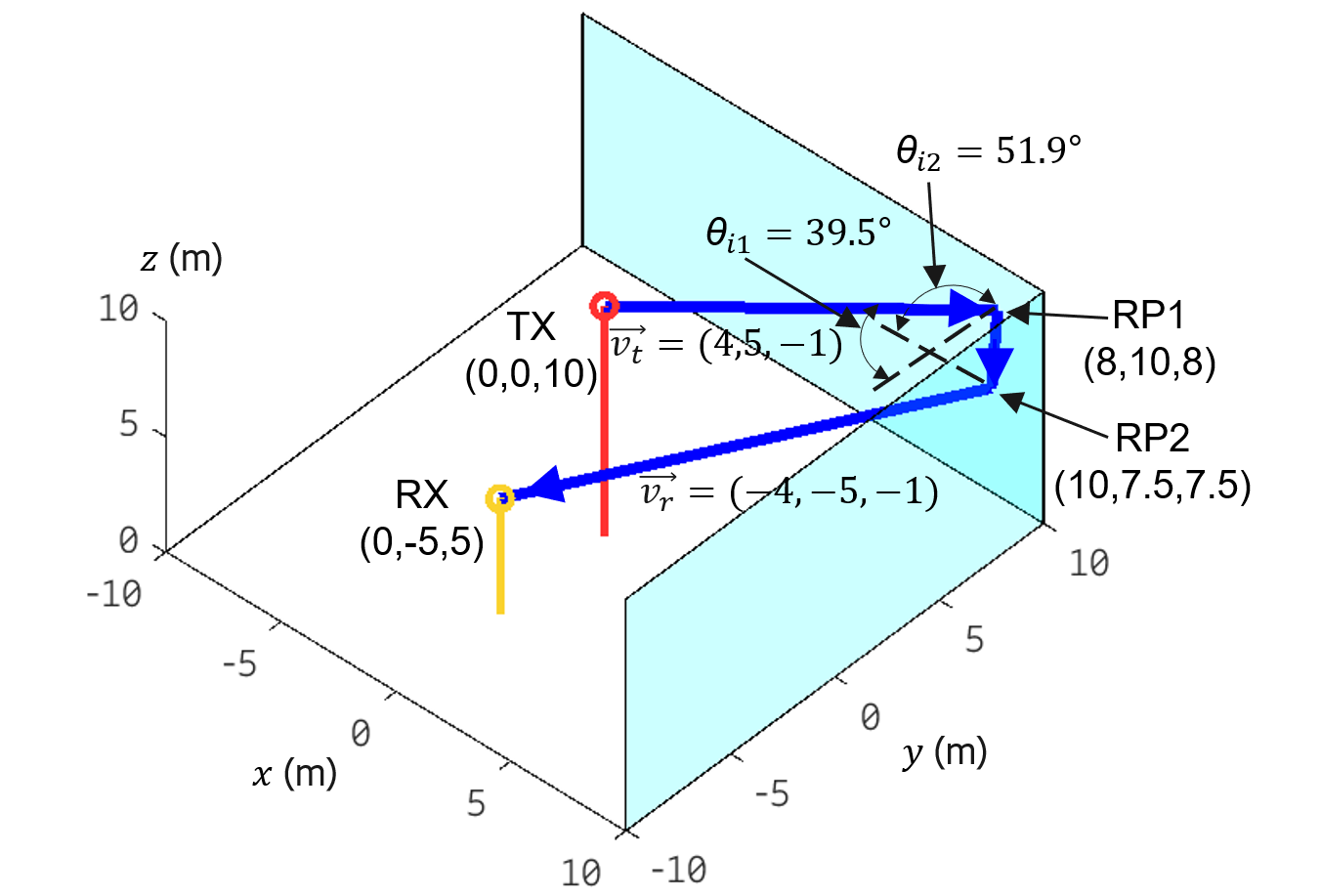}}\quad
	\subfloat[Trajectory B]{\label{fig:e}\includegraphics[width=0.8\columnwidth]{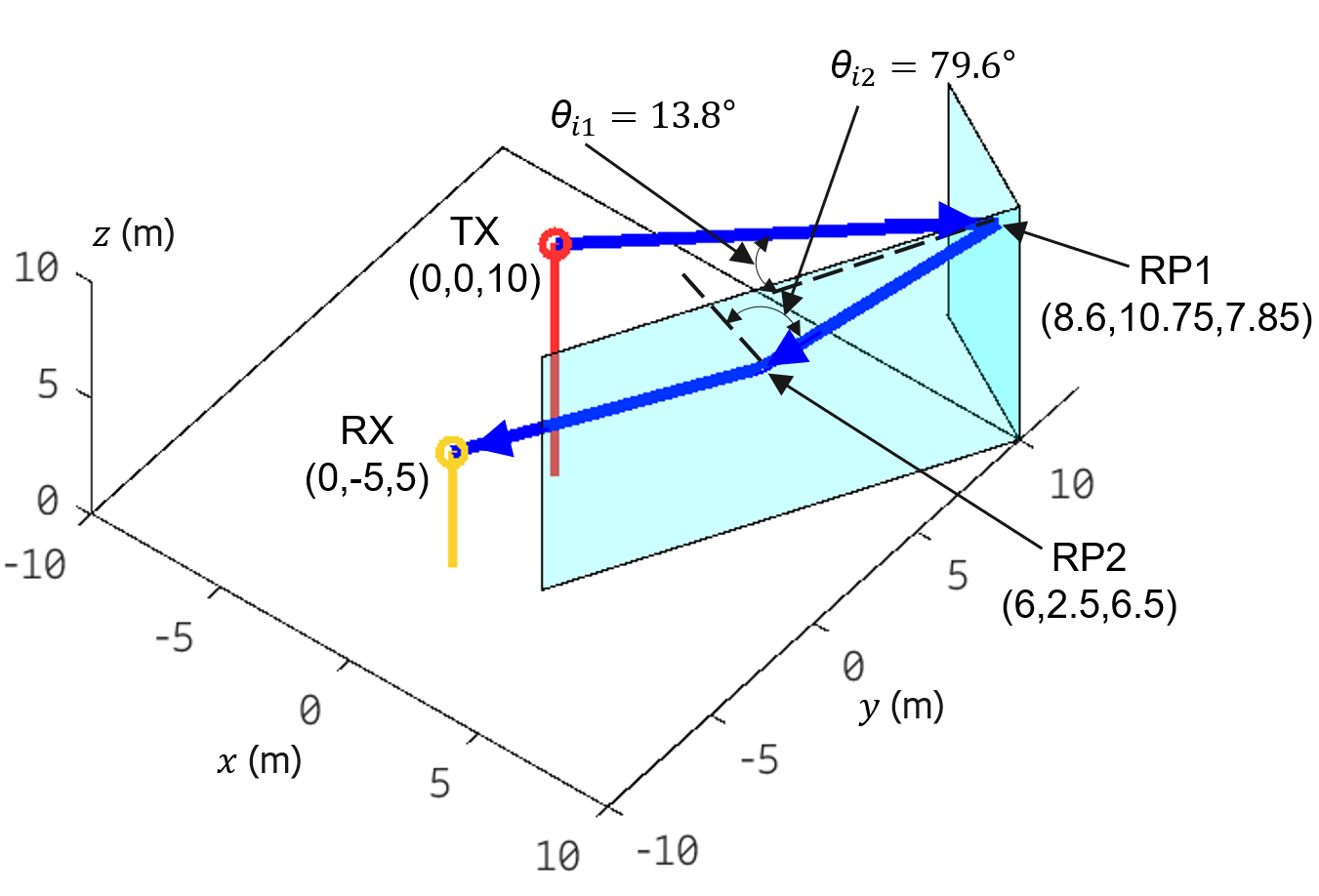}}\quad
	\subfloat[Trajectory C]{\label{fig:f}\includegraphics[width=0.8\columnwidth]{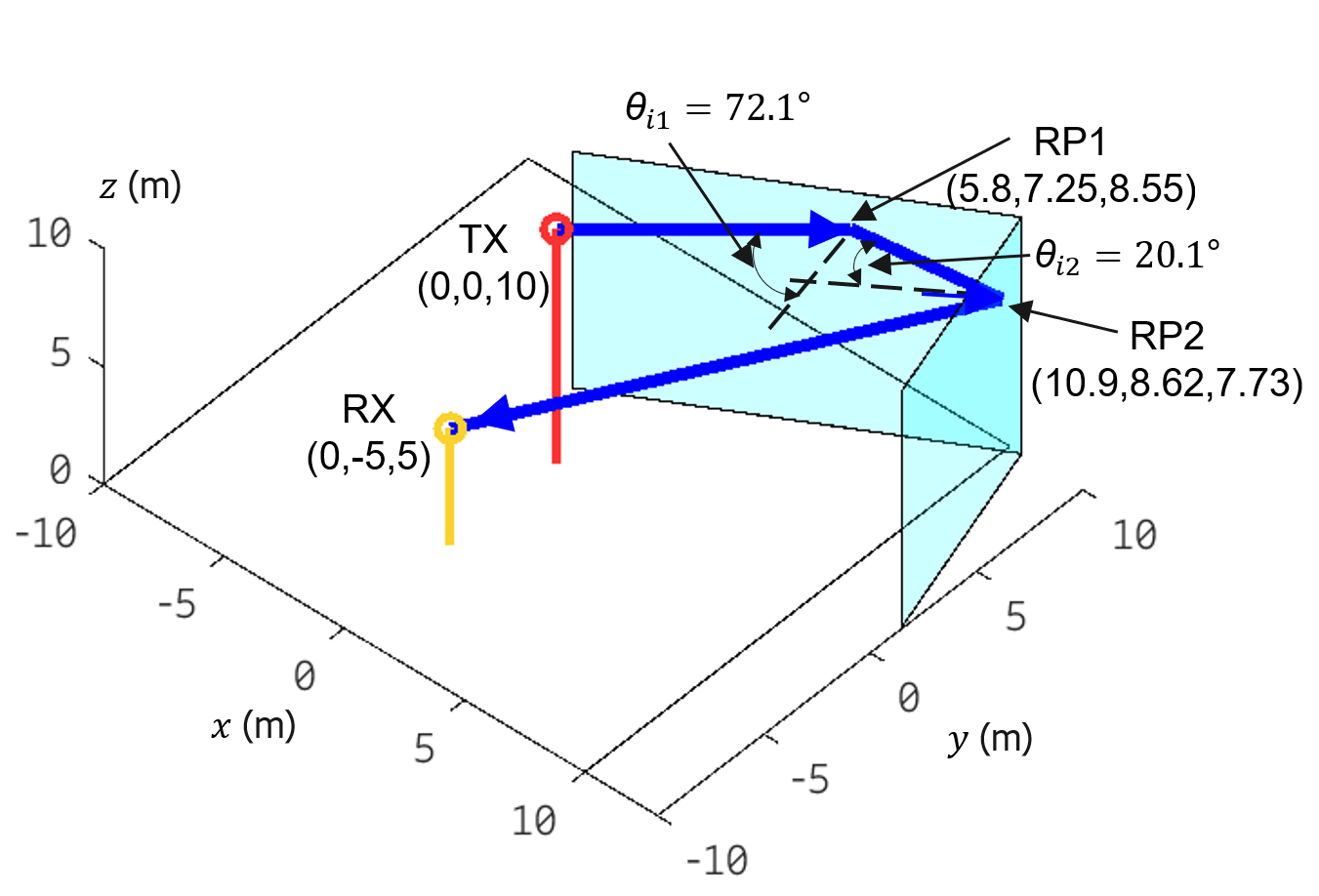}}\quad
	\subfloat[Trajectory D]{\label{fig:g}\includegraphics[width=0.8\columnwidth]{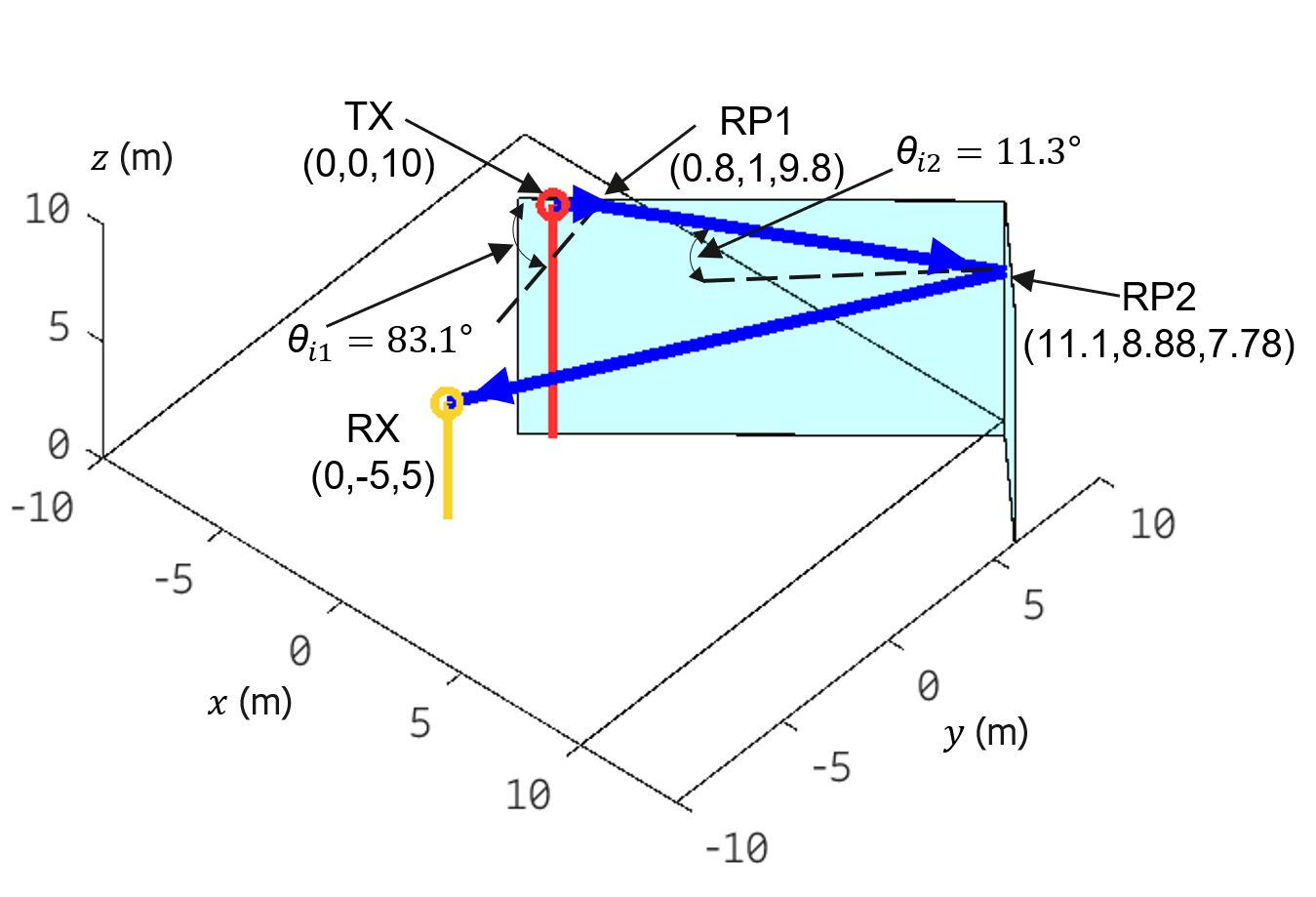}}\\	
	\caption{Four scenarios, which include four different scatterer layout, induce the same ray from \mbox{TX(-10, -5, 5)} in the direction of vector (4, 1, 0) to propagate along four different trajectories. However, from measurement point of view, all trajectories have same TX/RX locations, transmitter-side beam direction (4, 5, -1), receiver-side beam direction (-4, -5, -1), and overall path length of 32.4~m.}
	\label{fig_2}
\end{figure*}

\section{Challenges of RAT-Based Scatterer Detection and Localization}
Detecting and localizing scatterers surrounding radio transceivers in rich scattering environment is one of the emerging requirements of JCAS system. Most scatterer localization technologies trace radio trajectories and localize objects using 3D digital maps, ray tracing methods, and geometrical optics theory \cite{b3}. These map-based localization methods can be implemented only when the corresponding maps of the environment are available to the localization systems \cite{b4}. Therefore, map-based scatterer localization methods not only use cellular infrastructure but also depend on imported 3D map. However, 3D map is not the component of conventional cellular networks. Moreover, environmental changes are likely to happen frequently especially for indoor scenarios and lead to out-of-date information in the 3D digital maps. The objective of RAT-based scatterer localization is to sense and localize scatterers without requiring any external assistance beyond the scope of RAT capabilities, e.g., without importing the 3D digital map of an environment. With this restriction, tracing radio trajectory in rich scattering environment is quite challenging, because radio signals can be reflected by scatterers at unknown positions and attenuated by uncertain number of reflections. Possible information can be gleaned by RAT-based scatterer localization system including:

\begin{itemize}
	\item Angle of departure (AOD);
	\item Angle of arrival (AOA);
	\item Time of flight (TOF);
	\item Transmit power and received signal strength~(RSS);
	\item Positions of transmitters and receivers (anchor nodes).
\end{itemize}   

AOD and AOA represent transmitter- and receiver-side beam direction, respectively. TOF can be used to estimate the total path length. RSS at receiver side can be measured or calculated by link budget. Anchor nodes are transmitter or receiver with known coordinates \cite{b5}.

In rich scattering environment, non-line-of-sight (NLOS) radio signals typically undergo phenomena such as reflection, diffraction, and refraction. In this article, we only consider reflection for localization purpose. The reflection of NLOS trajectories can be either single- or multiple-bounce reflections. In prior works, the scatterer localization was done in a single-bounce context (single-bounce-assumption) \cite{b6}. Multiple-bounce reflection trajectories are excluded in most prior technologies to reduce the complexity of the localization algorithms. For example, typical scatterer localization system “radar” detects scatterers by exploiting single-bounce reflection. The sparse scatterers (e.g., planes and missiles) in the air make single-bounce-assumption reasonable for radar applications. Light detection and ranging (LiDAR) localizes scatterers with laser by measuring the time for the single order reflection between the transceivers. However, such approaches may not be applicable to RAT-based scatterer localization systems deployed in rich scattering environments, in which a large number of multiple-bounce reflection paths exist, especially in urban areas or cluttered indoor scenarios. Radio signals in such environments are likely to be reflected multiple times. Therefore, single-bounce reflection may not be sufficient to characterize the sensing parameters of scatterers.

With single-bounce-assumption, the sensing reliability and accuracy of RAT-based scatterer localization degrade in rich scattering environment. Scatterers may be mistakenly localized when radio rays are reflected between multiple scatterers. For example, as shown in Fig.~1, a TX antenna and a RX antenna are placed at \mbox{(-10, -5, 5)} and \mbox{(-10, 5, 5)} in a Cartesian coordinate system, respectively. The TX transmits a narrow beam in the direction of vector \mbox{$\overrightarrow {v_t}$=(4, 1, 0)}, the RX receives the beam in the direction of vector \mbox{$\overrightarrow{v_r}$=(-4, 1, 0)}. For this scenario, according to single-bounce-assumption, the reflection point~(RP) should locate at \mbox{(10, 0, 5)}, the radio ray should be reflected by the plane \mbox{$x-10=0$} producing single-bounce reflection, the trajectory is described by point-to-point motion passing through the points \mbox{TX(-10, -5, 5)}, \mbox{RP(10, 0, 5)}, and \mbox{RX(-10, 5, 5)} as shown in Fig.~1(a). Note that the scatterer represented by the plane \mbox{$x-10=0$} should be a larger surface than the radius of the first Fresnel zone. In order to demonstrate the spatial relationship between the trajectory and the scatterers, the scatterer is plotted to an appreciable rectangle. However, the trajectory in Fig.~1(a) may not represent the true trajectory of the radio ray. For example, if two scatterers represented by the planes \mbox{$1.47x-y-2.5=0$} and \mbox{$-1.47x-y+2.5=0$} exist as shown in Fig.~1(b), then the radio ray is reflected twice by the two scatterers and propagates along the trajectory passing through the points \mbox{TX(-10, -5, 5)}, \mbox{RP1(0, -2.5, 5)}, \mbox{RP2(0, 2.5, 5)}, and \mbox{RX(-10, 5, 5)}. If two scatterers \mbox{$x-y=0$} and \mbox{$y-10=0$} exist as shown in Fig.~1(c), the radio ray is reflected three times and propagates along the points \mbox{TX(-10, -5, 5)}, \mbox{RP1(-3.33, -3.33, 5)}, \mbox{RP2(0, 10, 5)}, \mbox{RP3(2, 2, 5)}, and \mbox{RX(-10, 5, 5)}. Any RAT-based scatterer localization system based on single-bounce-assumption cannot distinguish the trajectories in \mbox{Fig.~1(a)-(c)} due to exactly the same TX/RX locations, AOD, and AOA from the transceivers' point of view. In rich scattering environments (e.g., the urban canyons or the indoor scenarios), there are abundant scatterers in and around the propagation paths, such multiple-bounce-dominant paths are mistakenly identified as single-bounce reflection by the single-bounce-assumption-based localization methods. For example, as shown in Fig.~1(a) and Fig.~1(b), we assume that the sensors of an AD vehicle locate at \mbox{(-10, -5, 5)} and \mbox{(-10, 5, 5)}, and two close obstacles (e.g., pedestrians or other vehicles) locate at \mbox{(0, -2.5, 5)} and \mbox{(0, 2.5, 5)}. A single-bounce-assumption-based scatterer localization system cannot identify the two obstacles correctly: the two obstacles are identified as a single obstacle with wrong location \mbox{(10, 0, 5)} in the distance. This error may lead to wrong driving decision and fatal accident for AD.

\begin{figure*}[t]
	\centering
	\subfloat[Step 1 - simulate the PL of a single-bounce reflection trajectory reflection]{\label{fig:h}\includegraphics[width=0.65\columnwidth]{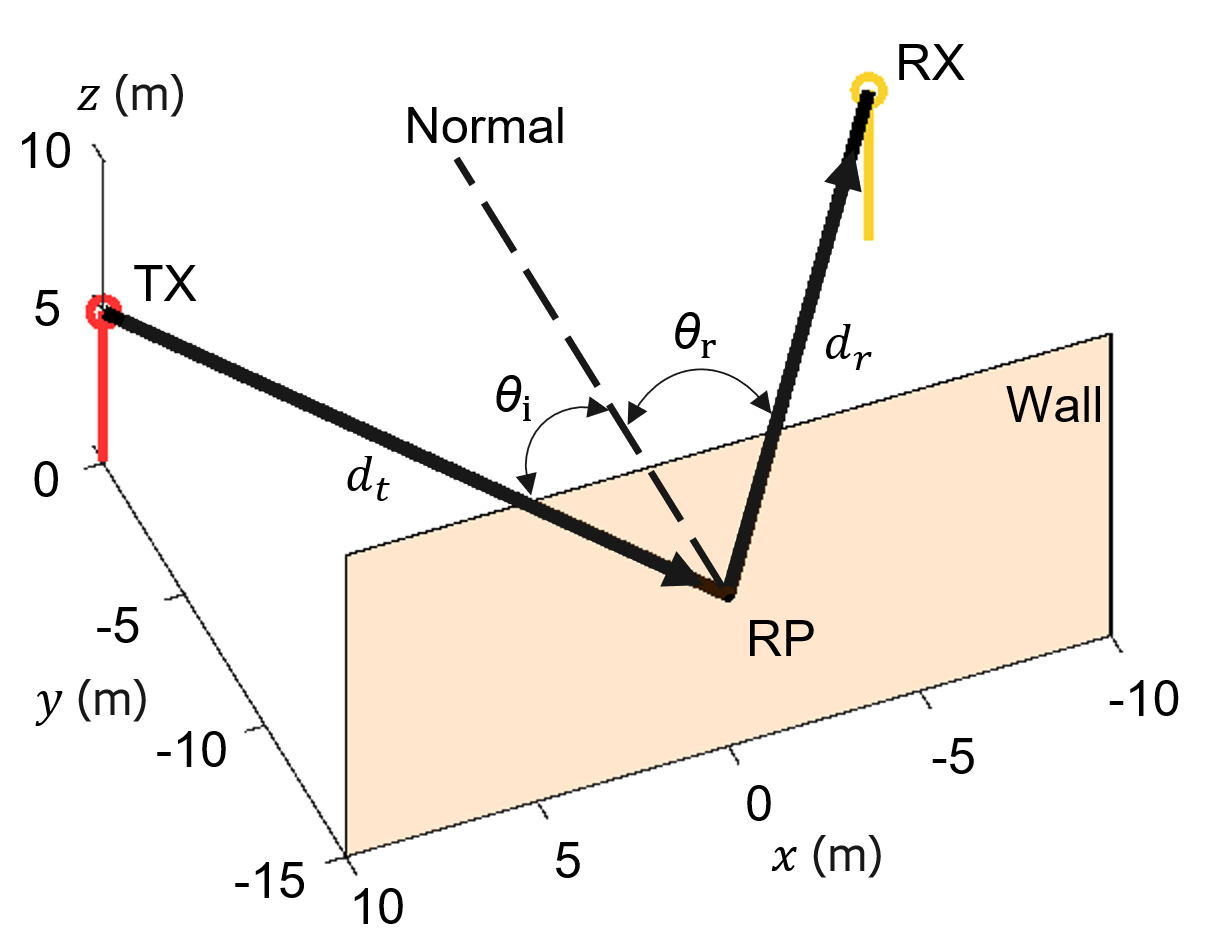}}\quad
	\subfloat[Step 1 - simulate the PL of a double-bounce reflection trajectory]{\label{fig:i}\includegraphics[width=0.6\columnwidth]{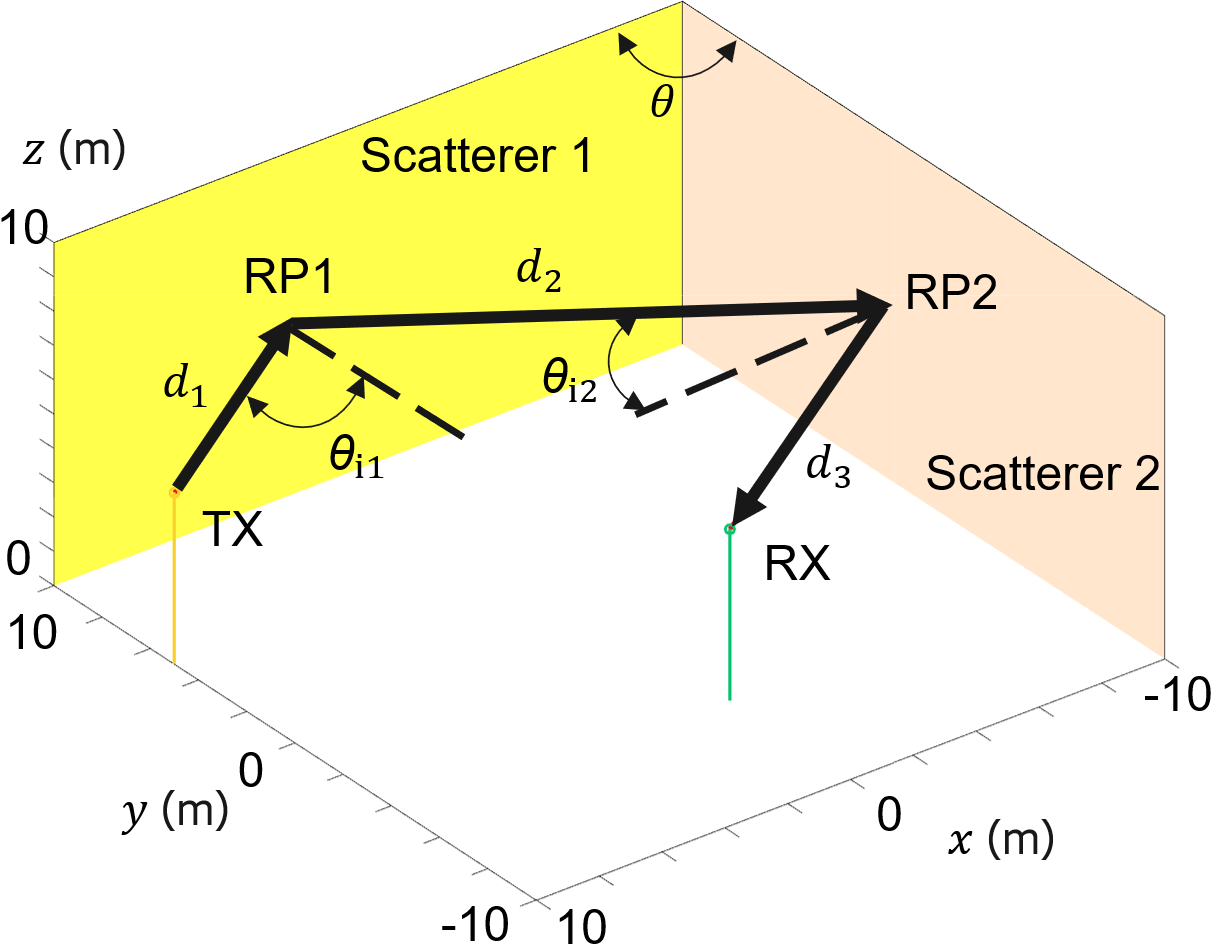}}\quad
	\subfloat[Step 2 - simulate the FSPL of a LOS path]{\label{fig:g}\includegraphics[width=0.65\columnwidth]{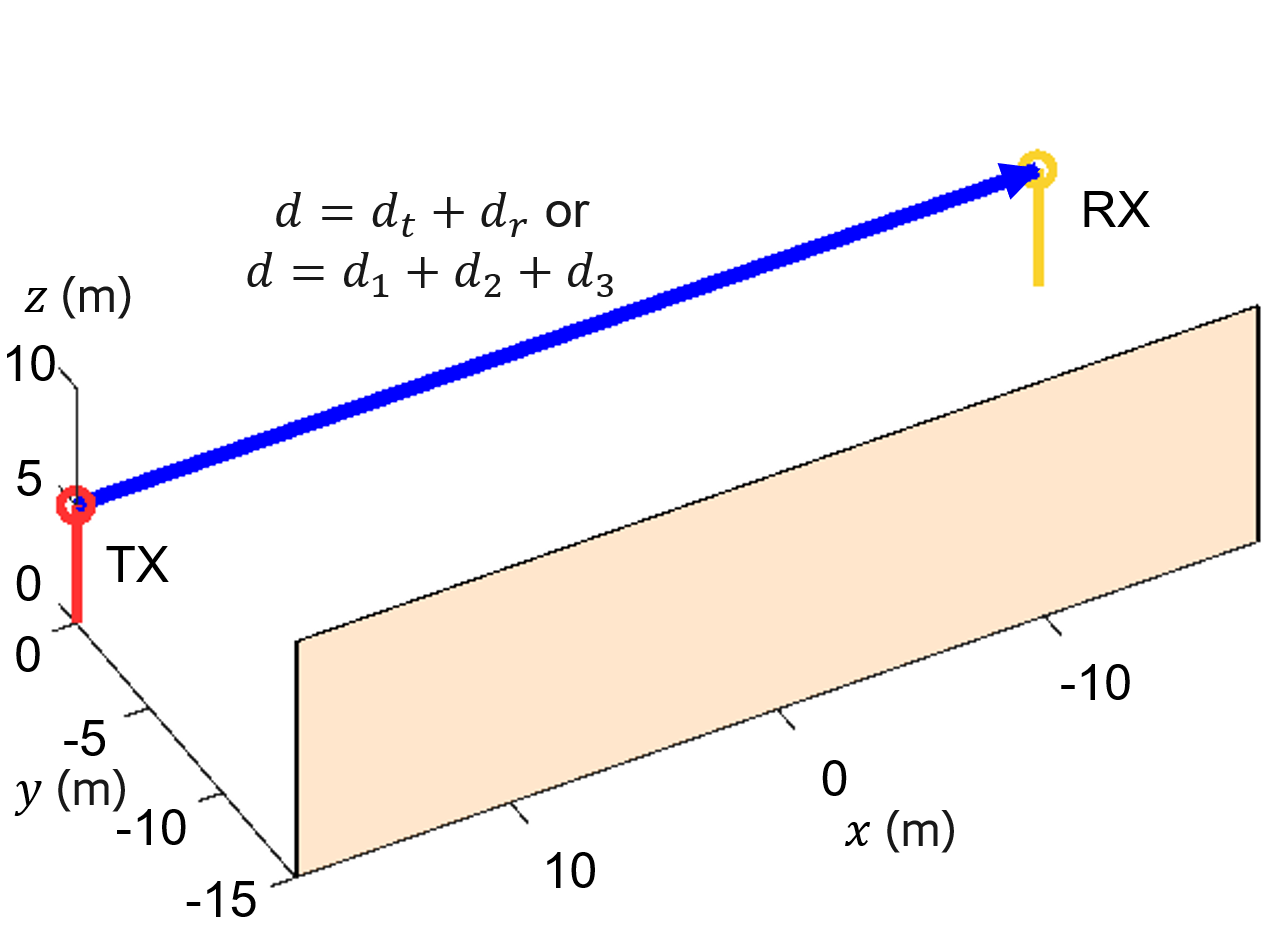}}\\	
	\caption{Matlab-based simulation steps to get the RL of single- and double-bounce reflection.}
	\label{fig_3}
\end{figure*}

\begin{figure*}[t]
	\centering
	\subfloat[RL of single-bounce reflection induced by wood, plasterboard, and glass at 100 GHz ]{\label{fig:a}\includegraphics[width=0.95\columnwidth]{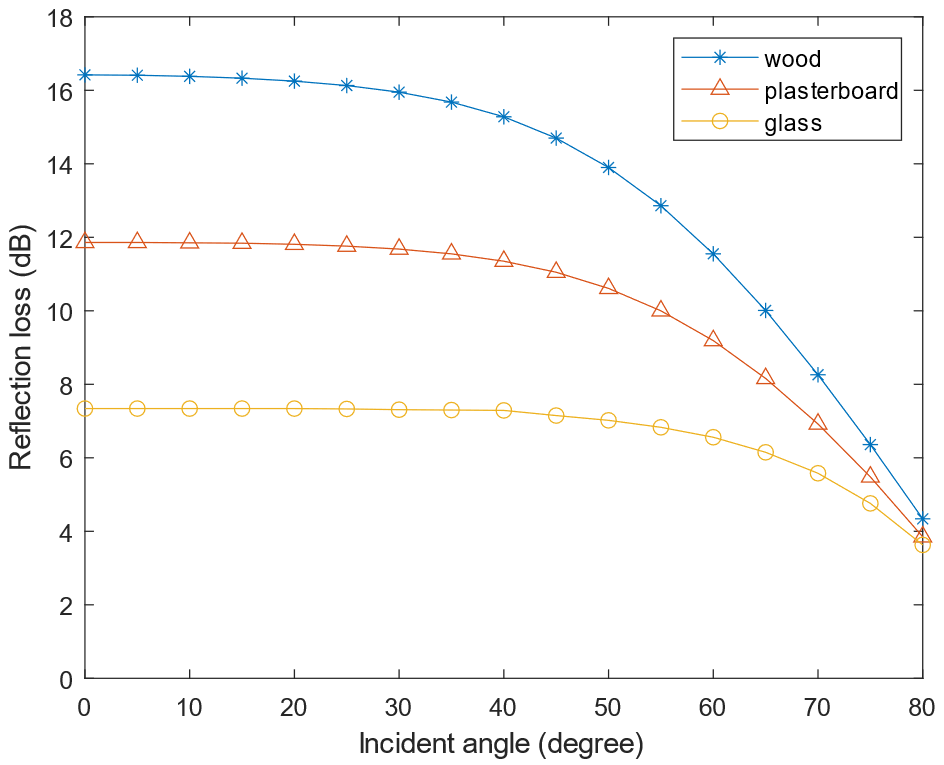}}\quad
	\subfloat[$\Sigma{RL}$ of double-bounce reflection induced by different sequences-of-material at any incident angles $\theta_{i1}$, $\theta_{i2}$ at 100 GHz ]{\label{fig:c}\includegraphics[width=1.05\columnwidth]{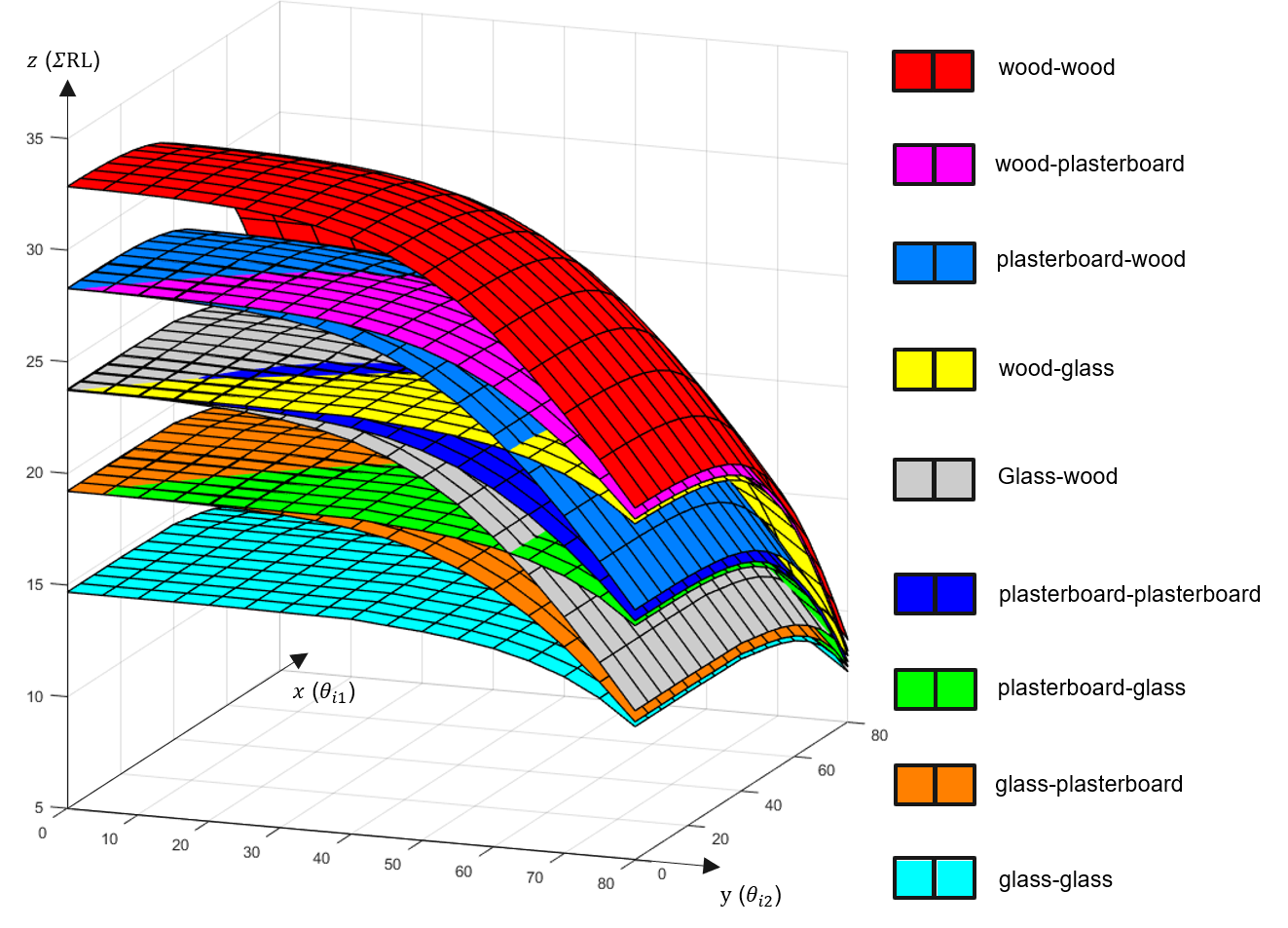}}\\	
	\caption{Simulated RL of single-bounce reflection and simulated $\Sigma$RL of double-bounce reflection at 100~GHz, and the scatterer/scatterers are made of any sequence of wood, plasterboard, and glass.}
\end{figure*}

Even if we assume that a RAT-based scatterer localization system knows the number of bounces of the radio rays, we still cannot precisely trace the trajectories and localize the scatterers. For example, as shown in Fig.~2, the transmitter- and receiver-side beam direction of a radio ray between TX\mbox{(0, 0, 10)} and RX\mbox{(0, -5, 5)} are \mbox{$\overrightarrow{v_t}$=(4, 5, -1)} and \mbox{$\overrightarrow{v_r}$=(-4, -5, -1)}, respectively. We assume that a RAT-based scatterer localization system knows the ray is reflected twice by the two surfaces of a dihedral, TOF is measured and the overall path length is 32.4~m. Different scatterers’ location and orientation illustrated in \mbox{Fig.~2(a)-(d)} induce four different trajectories (the trajectories from TX to RX with two reflections are represented as the concatenation of four coordinates separated by the hyphens): 

\begin{itemize}
	\item Trajectory A: 
	
	(0, 0, 10)-(8, 10, 8)-(10, 7.5, 7.5)-(0, -5, 5);
	\item Trajectory B:
	
	(0, 0, 10)-(8.6, 10.75, 7.85)-(6, 2.5, 6.5)-(0, -5, 5); 
	\item Trajectory C:
	
	(0, 0, 10)-(5.8, 7.25, 8.55)-(10.9, 8.62, 7.73)-(0, -5, 5); 
	\item Trajectory D:
	
	(0, 0, 10)-(0.8, 1, 9.8)-(11.1, 8.88, 7.78)-(0, -5, 5).
\end{itemize} 

It is difficult to distinguish the true trajectory from trajectories A-D by conventional scatterer localization methods because trajectories A-D are characterized by same TX/RX locations, same transmitter- and receiver-side beam direction, and same overall path length of 32.4~m.

\section{RL Numerical Validation}
\subsection{Simulations}
\begin{table}[t]
	\caption{Parameters of three common building materials recommended by ITU-R P.2040-1}
	\begin{center}
		\begin{tabular}{ccccccc}
			\hline
			\multirow{2}{*}{\textbf{Material}} &\multicolumn{2}{c}{\textbf{Permittivity}}&\multicolumn{2}{c}{\textbf{Conductivity}} \\
			& $a$ & $b$ & $c$ & $d$\\
			\hline
			Wood &	1.99 & 0 & 0.0047 & 1.0718 \\
			Plasterboard & 2.94 &	0 &	0.0116 & 0.7076\\
			Glass & 6.27 & 0 & 0.0043 & 1.1925 &\\
			\hline
		\end{tabular}
	\end{center}
	\label{tab_1}
\end{table}

To study the impact of RL on scatterer localization, we have carried out extensive RL simulations based on \mbox{MATLAB}. The simulation configuration is shown in Fig.~3, a TX antenna and a RX antenna are placed in an environment with the scatterers. By transmitting a highly directional radio ray from the TX, a single- and a double-bounce reflection trajectory between the TX and the RX can be simulated as shown in Fig.~3(a) and Fig.~3(b), respectively. The vertical walls can be made of one of the three common building materials in indoor environments, namely wood, plasterboard, and glass. The materials are homogeneous and the parameters of the materials in the simulation recommended by International Telecommunications Union (ITU) \cite{b7} are tabulated in Table~I. For single-bounce reflection, a radio ray transmitted by the TX strikes the vertical wall with an incident angle $\theta_i$ as shown in Fig.~3(a) \cite{b8}, $\theta_i$ is swept from 0\textdegree~to 80\textdegree~by changing the position of the RX. $d_t$ and $d_r$ are the path lengths between the RP and the TX/RX, respectively. $d$ is the total path length and $d=d_t+d_r$. For double-bounce reflection, a radio ray transmitted by the TX is reflected by the two surfaces of the dihedral at incident angles $\theta_{i1}$ and $\theta_{i2}$ as illustrated in Fig.~3(b). According to ray tracing and geometry theory \cite{b2}, the angle between two dihedral surfaces $\theta$ is equal to $\theta_{i1}$ plus $\theta_{i2}$. By steering transmitter-side beam direction and changing dihedral angle $\theta$, the combinations of ($\theta_{i1}, \theta_{i2}$) with any $\theta_{i1}$ and $\theta_{i2}$ value can be simulated. Let $d_1$, $d_2$, and $d_3$ be the distances between the TX and the RP1, the RP1 and the RP2, and the RP2 and the RX, respectively. Then the total path length $d$ of this double-bounce trajectory is equal to $d_1+d_2+d_3$.

By the simulation steps proposed in \cite{b8}, RL of single- and double-bounce reflection can be obtained. The received power at the RX side, which is attenuated by free space propagation and reflection at a certain incident angle, is measured in the first step. The overall path loss (PL) in decibel including RL and free space path loss (FSPL) of the trajectory is calculated as:
\begin{equation}\label{eqn_1}
PL = P_{TX} - P_{RX},
\end{equation}
where $P_{TX}$ is the transmit power in dBm, $P_{RX}$ is the received power in dBm.

In the second step, the RX is placed at the distance $d$ ($d=d_t+d_r$ for single-bounce or $d=d_1+d_2+d_3$ for double-bounce) from the TX. The TX transmits exactly the same radio ray as the one in step~1. Therefore, a LOS trajectory between the TX and the RX is established as illustrated in Fig.~3(c). The FSPL at a distance of $d$ can be measured, or can be calculated by Friis’ equation:
 \begin{equation}\label{eqn_2}
FSPL(f,d) = 32.4 +20\log_{10}(f) + 20\log_{10}(d),
\end{equation}
where $f$ is the carrier frequency in mega Hertz and $d$ is path length in kilometer.

Finally, the RL can be calculated by subtracting \eqref{eqn_2} from \eqref{eqn_1}:
\begin{equation}\label{eqn_3}
RL = PL - FSPL(f,d).
\end{equation}

It is worth noting that RL takes no account of loss in free space prior or subsequent to the interaction of a wave with the scatterer \cite{b7}, RL depends on the interaction between the wave and the scatterer only. Antenna type, antenna gain, transmit power, and path length do not affect RL. A change in these factors brings a corresponding change in RSS. For example, if the $P_{TX}$ increases by 3~dB, the measured RSS increases by 3~dB accordingly. From \eqref{eqn_1}-\eqref{eqn_3}, the increased transmit power has no impact on PL and RL at all.

The RL simulation results of single-bounce reflection induced by scatterers made of wood, plasterboard, and glass at different incident angle $\theta_i$ at 100~GHz are reported in Table~II and Fig.~4(a). The roughness of wood, plasterboard, and glass in the simulation are 0.4~mm, 0.2~mm, and 0~mm (perfectly smooth), respectively. The roughness represents the root mean square value of the height deviation from a perfectly smooth surface \cite{b9}. From Fig.~4(a), for scatterers made of specific material, RL and incident angle are negatively correlated, large incident angle induces small RL \cite{b8}.

The sum of reflection losses ($\Sigma$RL) simulation results of double-bounce reflection induced by two scatterers made of any combinations of wood, plasterboard, and glass at incident angles $\theta_{i1}$ and $\theta_{i2}$ at 100~GHz are reported in Fig.~4(b). In Fig.~4(b), \mbox{$x$-axis} represents the first incident angle $\theta_{i1}$ of a radio ray reflected by the first scatterer, \mbox{$y$-axis} represents the second incident angle $\theta_{i2}$ of same ray reflected by the second scatterer, \mbox{$z$-axis} represents the $\Sigma$RL (RL1 plus RL2) induced by the two scatterers. The collection of $\Sigma$RL data of each sequence-of-material is plotted to a surface and displayed with specific color as shown in Fig.~4(b). A sequence-of-material, which contains a set of materials, represents the material of scatterers in temporal order when the radio ray strike the scatterers in sequence. The legends in Fig.~4(b) show all sequences-of-material of double-bounce reflection. For example, the magenta surface in Fig.~4(b) represents the $\Sigma$RL induced by double-bounce reflection, and the first scatterer is made of wood, the second scatterer is made of plasterboard. From Fig.~4(b), it is evident that the $\Sigma$RLs induced by different sequences-of-material are significantly different, especially when both incident angles $\theta_{i1}$ and $\theta_{i2}$ are small. When both incident angles increase, a sharp reduction in $\Sigma$RL is observed. This is expected behavior due to the $\Sigma$RL of double-bounce reflection can be divided into two single-bounce reflections induced by two scatterers respectively, and each single-bounce reflection induces low RL at large incident angle as shown in Fig.~4(a).

From the simulation results, it is worth noting that the $\Sigma$RL induced by two scatterers is independent of scatterers’ sequence-of-material when a radio ray strikes the two scatterers. For example, a radio ray reflected by a woody surface at incident angle $\theta_{i1}$ of 50\textdegree~first and then reflected by a glassy surface at incident angle $\theta_{i2}$ of 75\textdegree~has same $\Sigma$RL of 18.66~dB as same radio ray reflected by glassy surface at incident angle $\theta_{i1}$ of 75\textdegree~first and then reflected by woody surface at incident angle $\theta_{i2}$ of 50\textdegree. Therefore, by $\Sigma$RL of double-bounce reflection, scatterers’ material combination may be concluded but sequence-of-material of two scatterers is still uncertain.

\subsection{Theory}
The RL simulation results in Fig.~4 can be derived from Fresnel equations \cite{b7} as follows. The relative permittivity of a material is given by
\begin{equation}\label{eqn_4}
	\eta = af^b -j17.98cf^d/f,
\end{equation}
where $\eta$ is the relative permittivity of the material; $a$, $b$, $c$, and $d$ are material properties that determine the relative permittivity and conductivity. The value of $a$, $b$, $c$, and $d$ are given in Table~I.

 Fresnel reflection coefficients $r_{TE}$ and $r_{TM}$ for transverse electric (TE) and transverse magnetic (TM) polarization describe the ratio of the amplitude of the reflected wave to the amplitude of the incident wave when the wave incidents upon a material at a certain frequency and at incident angle $\theta_i$ from the air \cite{b7}:
\begin{equation}\label{eqn_5}
	r_{TE} = \frac{cos{\theta_i}-\sqrt{\eta-sin^2{\theta_i}}}{cos{\theta_i}+\sqrt{\eta-sin^2{\theta_i}}},
\end{equation}

\begin{equation}\label{eqn_6}
	r_{TM} = \frac{{\eta}cos{\theta_i}-\sqrt{\eta-sin^2{\theta_i}}}{{\eta}cos{\theta_i}+\sqrt{\eta-sin^2{\theta_i}}}.
\end{equation}

As explained above, RL is dependent on material of scatterer, radio frequency, and incident angle only. This conclusion is also well supported by Fresnel equations \eqref{eqn_4}-\eqref{eqn_6}. From \eqref{eqn_4}-\eqref{eqn_6}, it is clear that reflection coefficients take no account of free space losses, but only the effect of the media interface \cite{b7}. Antenna type, antenna gain, transmit power, and path length do not contribute to RL at all.

Fig.~5 gives the reflection coefficient amplitude for a wave at 100~GHz in the air incident upon wood, plasterboard, and glass over a range of incident angles from 0° to 90° for both TE and TM polarization using \eqref{eqn_4} to \eqref{eqn_6} and taking the properties of materials in Table~I. But in practice we are more interested in formulae that determine reflection coefficient power, since power of radio signals can be directly measured by radio transceivers. The power of a wave is generally proportional to the square of the wave’s amplitude. Therefore, Fresnel reflection coefficient power $R_{TE}$ and $R_{TM}$ are the square of $r_{TE}$ and $r_{TM}$:
\begin{equation}\label{eqn_7}
	R_{TE} = |r_{TE}|^2,
\end{equation}

\begin{equation}\label{eqn_8}
	R_{TM} = |r_{TM}|^2.
\end{equation}

In practical applications, effective power coefficient $R_e$, which is the average of $R_{TE}$ and $R_{TM}$, is reasonable for cross-polarized antennas commonly used by cellular networks:

\begin{equation}\label{eqn_9}
	R_e = \frac{1}{2}(R_{TE}+R_{TM}).
\end{equation}

Fig.~6 gives the reflection coefficient power for a wave at 100~GHz in the air incident upon wood, plasterboard, and glass over a range of incident angles from 0° to 90° for both TE and TM polarization using \eqref{eqn_4} to \eqref{eqn_9} and taking the properties of materials in Table~I.

Fresnel equations \eqref{eqn_4}-\eqref{eqn_9} can be applied to perfectly smooth surfaces only. In order to include scattering loss induced by rough surface, the Fresnel reflection coefficients $r_{TE}$ and $r_{TM}$ should be multiplied by the Rayleigh roughness factor $\rho$ \cite{b10}:

\begin{equation}\label{eqn_10}
	{r'_{TE}}=\rho\cdot{r_{TE}},
\end{equation}

\begin{equation}\label{eqn_11}
	{r'_{TM}}=\rho\cdot{r_{TM}},
\end{equation}
where $r'_{TE}$ and $r'_{TM}$ are the modified reflection coefficients with the impact of roughness for TE and TM polarization, respectively. The Rayleigh roughness factor $\rho$ is given by:

\begin{equation}\label{eqn_12}
	\rho = e^{-\frac{g}{2}},
\end{equation}
with
\begin{equation}\label{eqn_13}
	g = (\frac{4\pi{\sigma}_Rcos{\theta_i}}{\lambda})^2,
\end{equation}
where $g$ is a parameter established for the effect of roughness \cite{b11}, $\sigma_R$ is the standard deviation of the surface roughness and $\lambda$ is the wavelength of radio wave.

A summary of the RL calculation consists of the following steps. First, perform relative permittivity calculation by using \eqref{eqn_4} and taking the parameters of the materials from Table~I. Second, the TE and TM components of a radio wave can be separately calculated by using \eqref{eqn_5}-\eqref{eqn_8}, then they are re-combined by using \eqref{eqn_9}. The impact of roughness on RL can be involved by \eqref{eqn_10}-\eqref{eqn_13}. Finally, the RL can be calculated by converting the effective power coefficient $R_e$ in percentage terms into RL in decibel.

\begin{table*}[h]
	\centering
	\caption{An example of RL database of single-bounce reflection induced by wood, plasterboard, and glass at 100 GHz}\label{tab_2}
	\begin{tabular}{@{} l*{17}{>{$}c<{$}} @{}}
		\toprule
		\textbf{Material} & \multicolumn{17}{c@{}}{\textbf{RL in decibel at different incident angle in degree}}\\
		\cmidrule(l){2-17}
		& \mathrm{0^{o}} & \mathrm{5^{o}} &  \mathrm{10^{o}} &  \mathrm{15^{o}} &  \mathrm{20^{o}} &  \mathrm{25^{o}} &  \mathrm{30^{o}} &  \mathrm{35^{o}} &  \mathrm{40^{o}} &  \mathrm{45^{o}} &  \mathrm{50^{o}} &  \mathrm{55^{o}} &  \mathrm{60^{o}} &  \mathrm{65^{o}} &  \mathrm{70^{o}} &  \mathrm{75^{o}} &  \mathrm{80^{o}} \\
		\midrule
		Wood & 16.42 & 16.41 & 16.38 & 16.33 & 16.25 & 16.13 & 15.95 & 15.68 & 15.28 & 14.7 & 13.9 & 12.86 & 11.55 & 10.01 & 8.26 & 6.36 & 4.34 \\[1ex]	

		Plasterboard & 11.86 & 11.86 & 11.85 & 11.84 & 11.81 & 11.76 & 11.68 & 11.55 & 11.35 & 11.05 & 10.61 & 10 & 9.19 & 8.16 & 6.92 & 5.48 & 3.85 \\[1ex]

		Glass & 7.34 & 7.34 & 7.34 & 7.34 & 7.34 & 7.33 & 7.31 & 7.3 & 7.29 & 7.15 & 7.02 & 6.83 & 6.56 & 6.15 & 5.58 & 4.76 & 3.63 \\[1ex]
		
	\end{tabular}
\end{table*}

\begin{figure*}[t]
	\centering
	\subfloat[Wood]{\label{fig:h}\includegraphics[width=0.65\columnwidth]{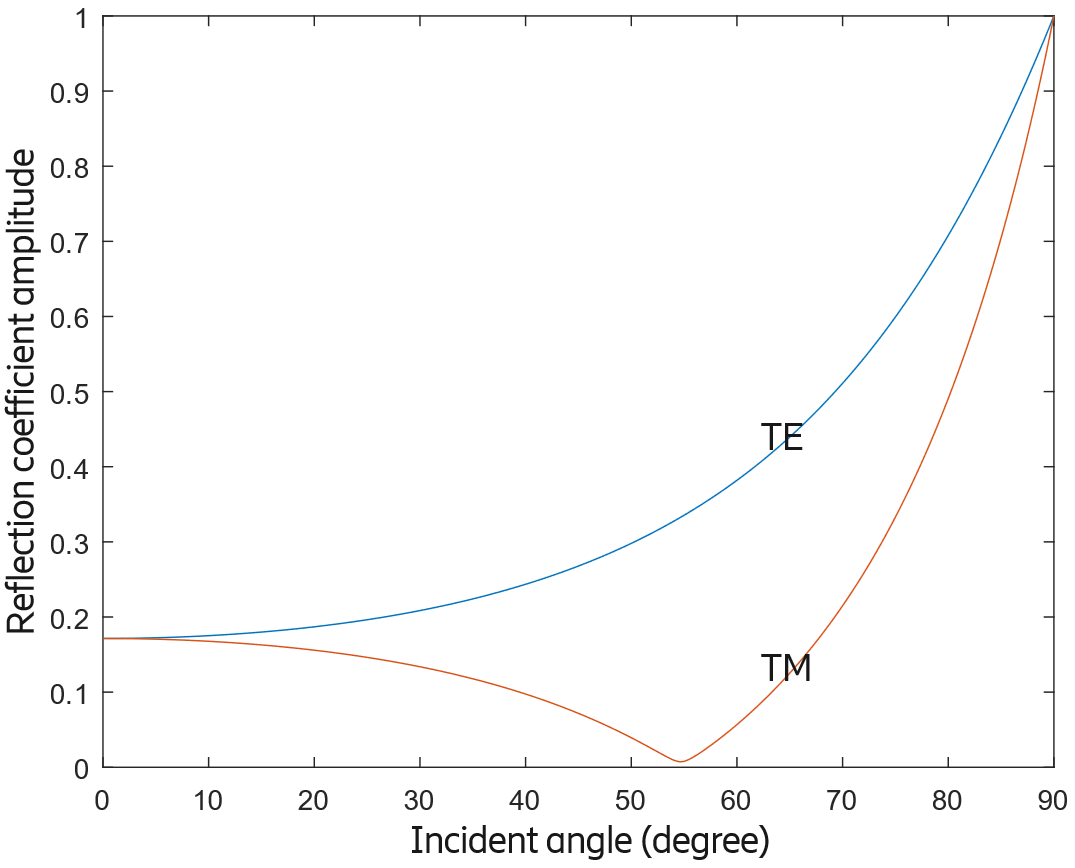}}\quad
	\subfloat[Plasterboard]{\label{fig:i}\includegraphics[width=0.65\columnwidth]{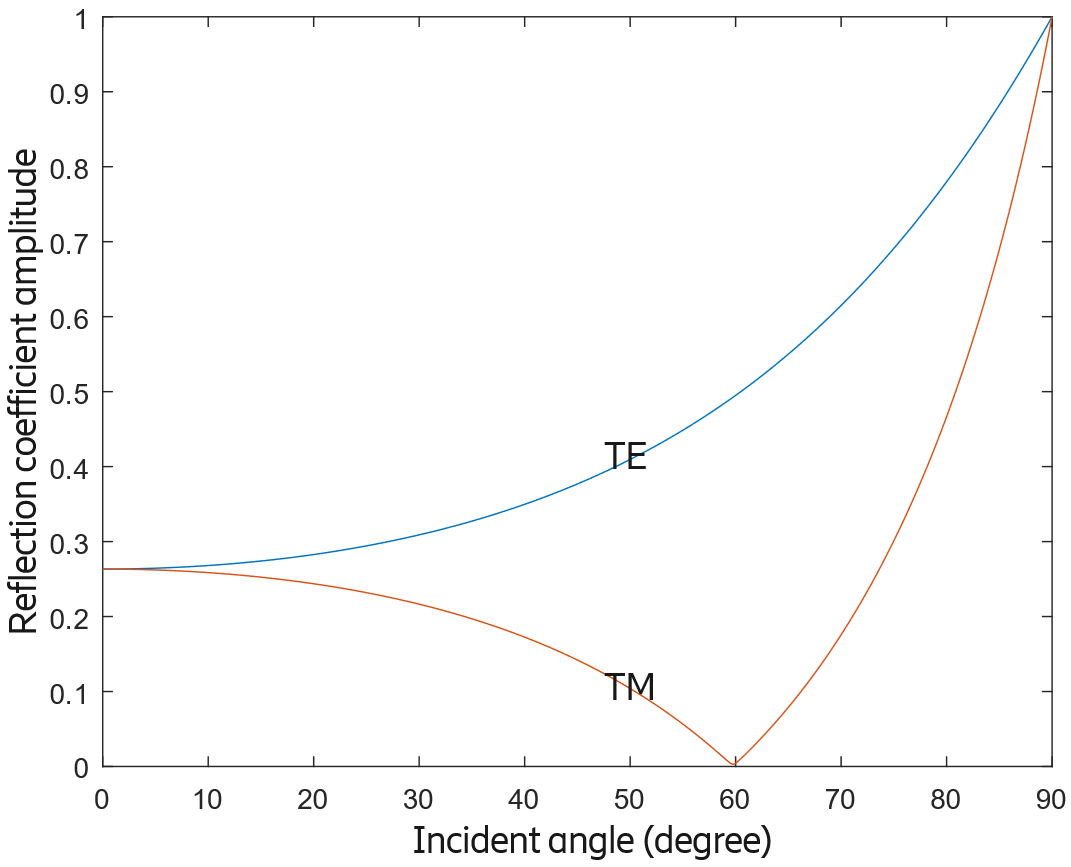}}\quad
	\subfloat[Glass]{\label{fig:g}\includegraphics[width=0.65\columnwidth]{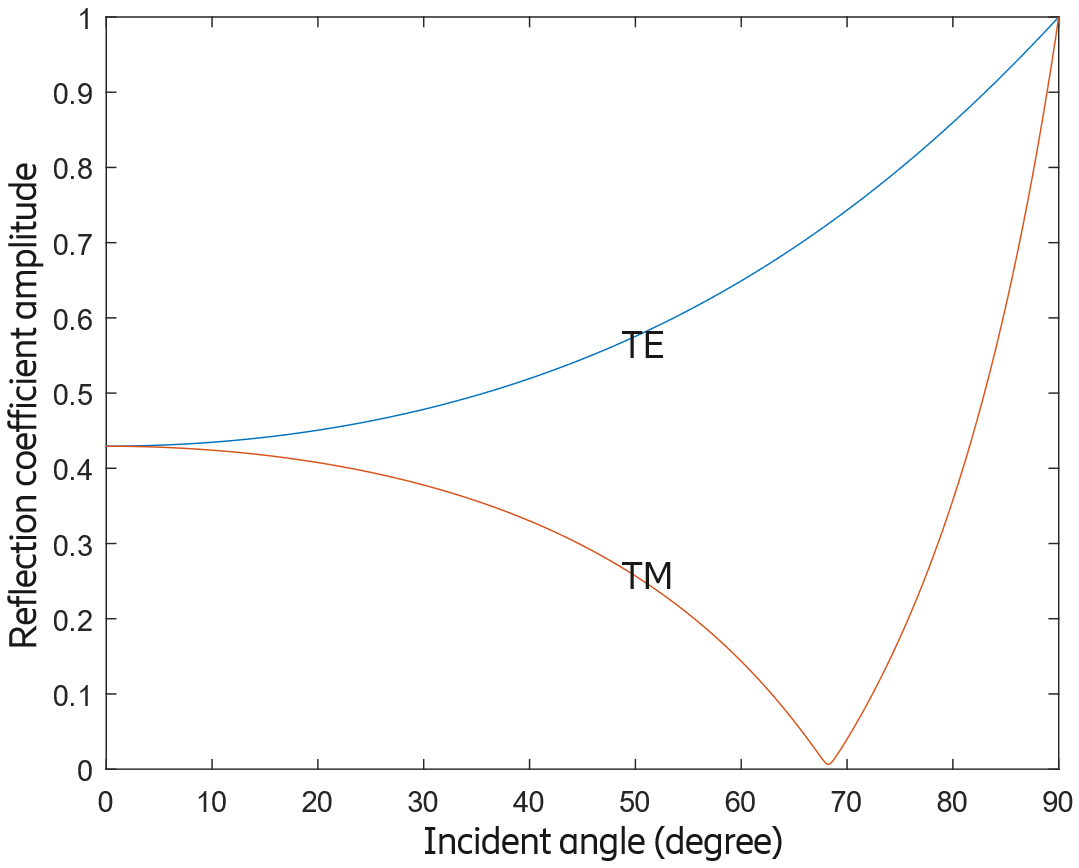}}\\	
	\caption{Reflection coefficient amplitudes for air/wood, air/plasterboard, and air/glass interfaces at 100~GHz.}
	\label{fig_5}
\end{figure*}

\begin{figure*}[t]
	\centering
	\subfloat[Wood]{\label{fig:h}\includegraphics[width=0.65\columnwidth]{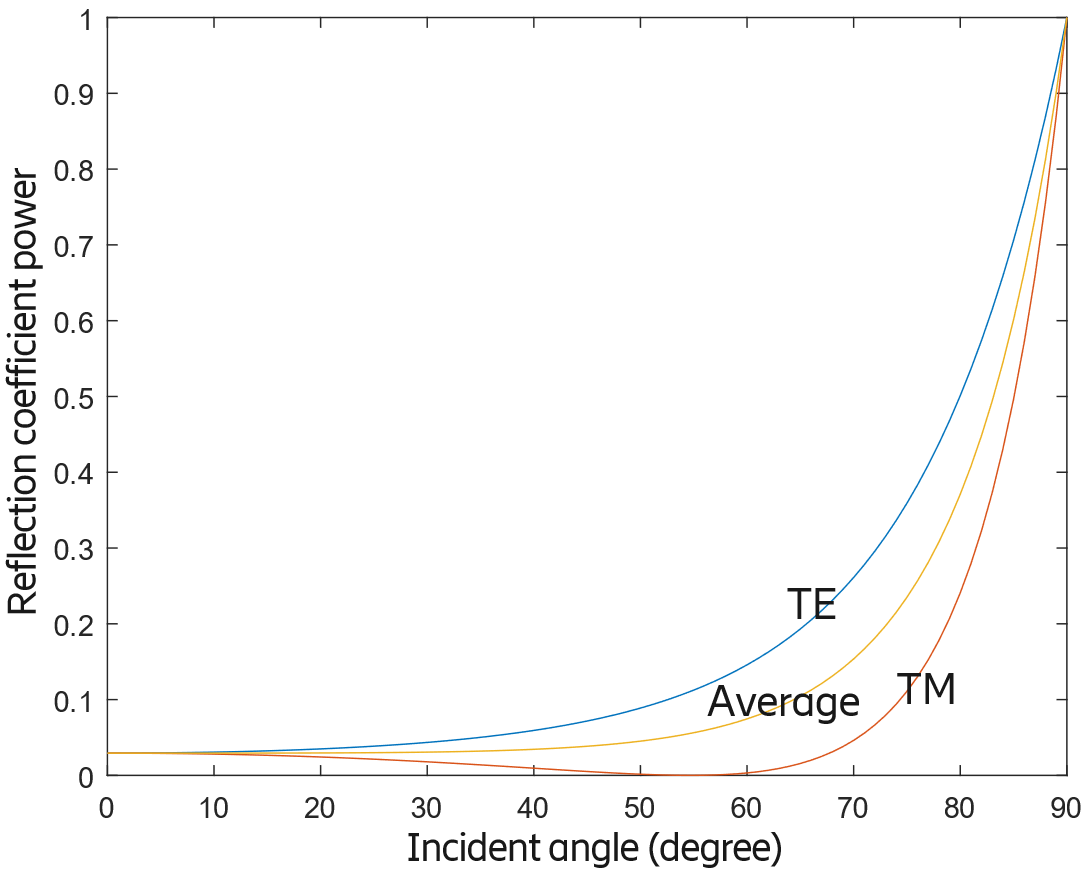}}\quad
	\subfloat[Plasterboard]{\label{fig:i}\includegraphics[width=0.65\columnwidth]{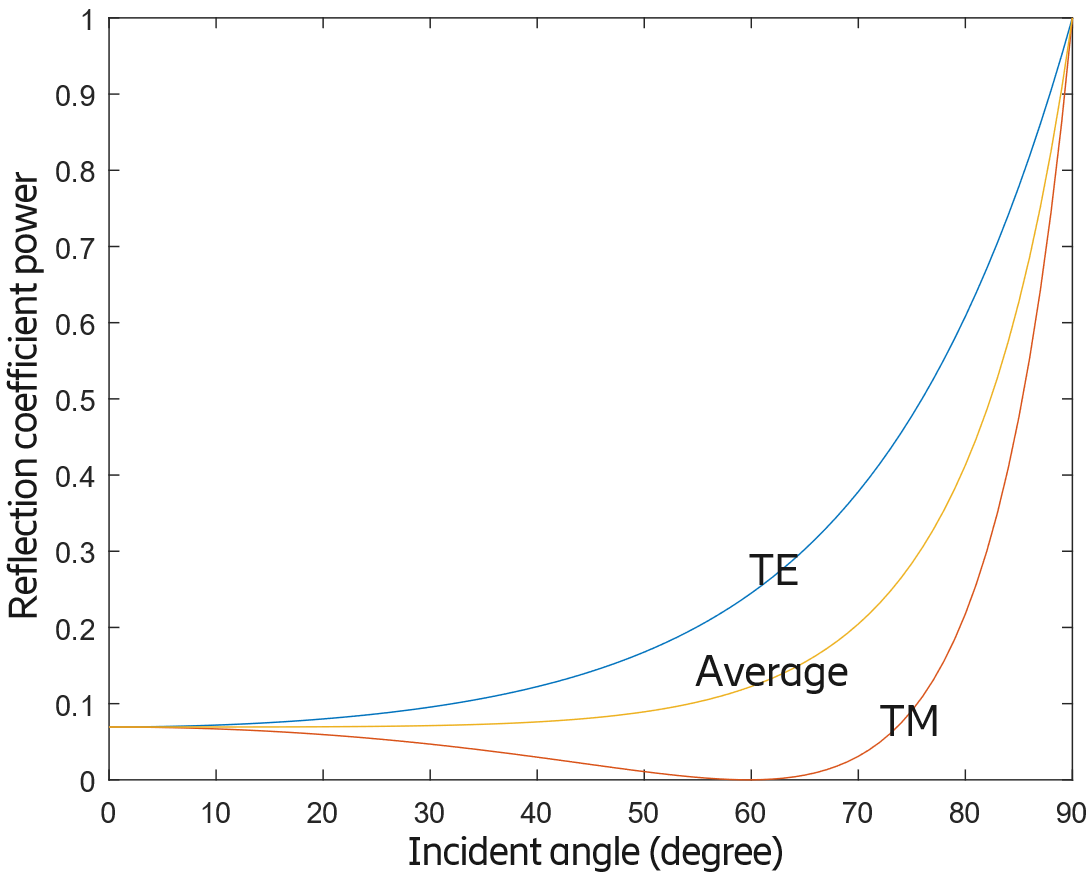}}\quad
	\subfloat[Glass]{\label{fig:g}\includegraphics[width=0.65\columnwidth]{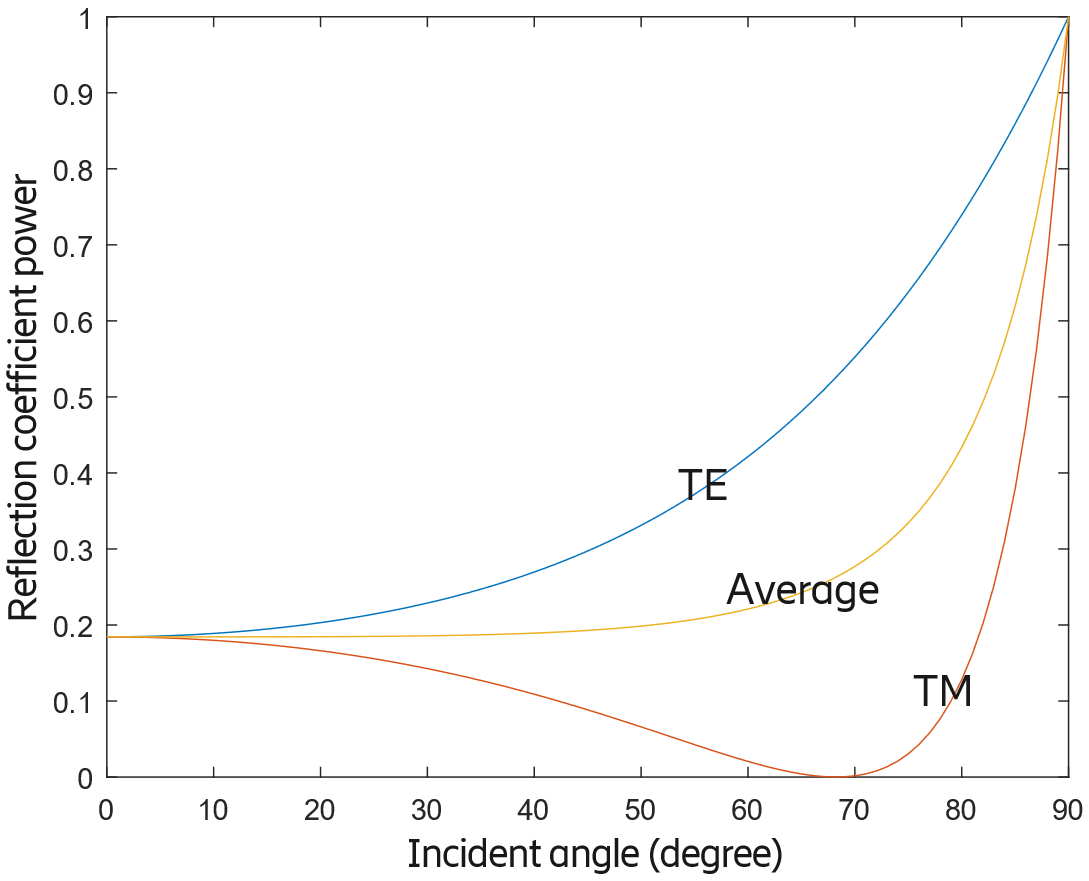}}\\	
	\caption{Reflection coefficient power for air/wood, air/plasterboard, and air/glass interfaces at 100~GHz.}
	\label{fig_6}
\end{figure*}

\section{RAT-Based Scatterer Localization and Material Identification Methods}
\subsection{Methods}
To tackle the challenges of RAT-based scatterer localization and material identification in rich scattering environment, we propose two novel methods all based on RL to localize scatterers and identify scatterers' materials simultaneously for NLOS trajectories. These methods do not require any prior knowledge of the environment. The first method (referred to as "method~1" below) comprises the following steps:

\begin{itemize}
	\item Step 1: calculate RL or $\Sigma$RL induced by scatterers under different scenarios at various frequencies to be used for scatterer localization. A specific scenario consists of the following information:
	the material and incident angle of the n-th scatterer (n=1,2,...,N). The scenario information (materials and incident angles), frequencies, and corresponding RLs are stored in a RL database.
	\item Step 2: establish a NLOS trajectory between TX and RX in an environment to be detected at a certain frequency and measuring the RL or $\Sigma$RL for this trajectory by using the measurement method proposed in Section~III.
	\item Step 3: identify a set of possible trajectories by matching the overall path length.
	\item Step 4: for the possible trajectories obtained by Step~3, identify the true trajectory by comparing the measured RL or $\Sigma$RL at a certain frequency with pre-calculated RL or $\Sigma$RL in the RL database created in Step~1.
\end{itemize} 

For the sake of simplicity, we assume that all scatterers in an environment are made of the materials listed in Table~I only, and the number of bounces of each trajectory is no more than two. Hence, the NLOS trajectories are divided into two categories: single-bounce reflection (N=1) with one RP and double-bounce reflection (N=2) with two RPs. For single-bounce reflection, the scatterer can be made of any of the three materials namely: wood, plasterboard, and glass. For double-bounce reflection, there are three possibilities of material for the first scatterer, three for the second scatterer. Therefore, there are nine possible sequences-of-material of the two scatterers in total. In Step~1, a RL data collecting phase is needed before the utilization of the proposed method. RLs induced by any sequences-of-material at any incident angles are calculated by Fresnel equations, then all the relevant information is stored in a RL database. An example of RL database of single-bounce reflection is shown in Table~II. By using the information in Table~II, the single-bounce RL induced by any material at any incident angle can be obtained. For multiple-bounce reflection, the $\Sigma$RL can be obtained by adding the RL of each reflection together.

\begin{figure}[t]
	\centerline{\includegraphics[width=\linewidth, height=10cm, keepaspectratio]{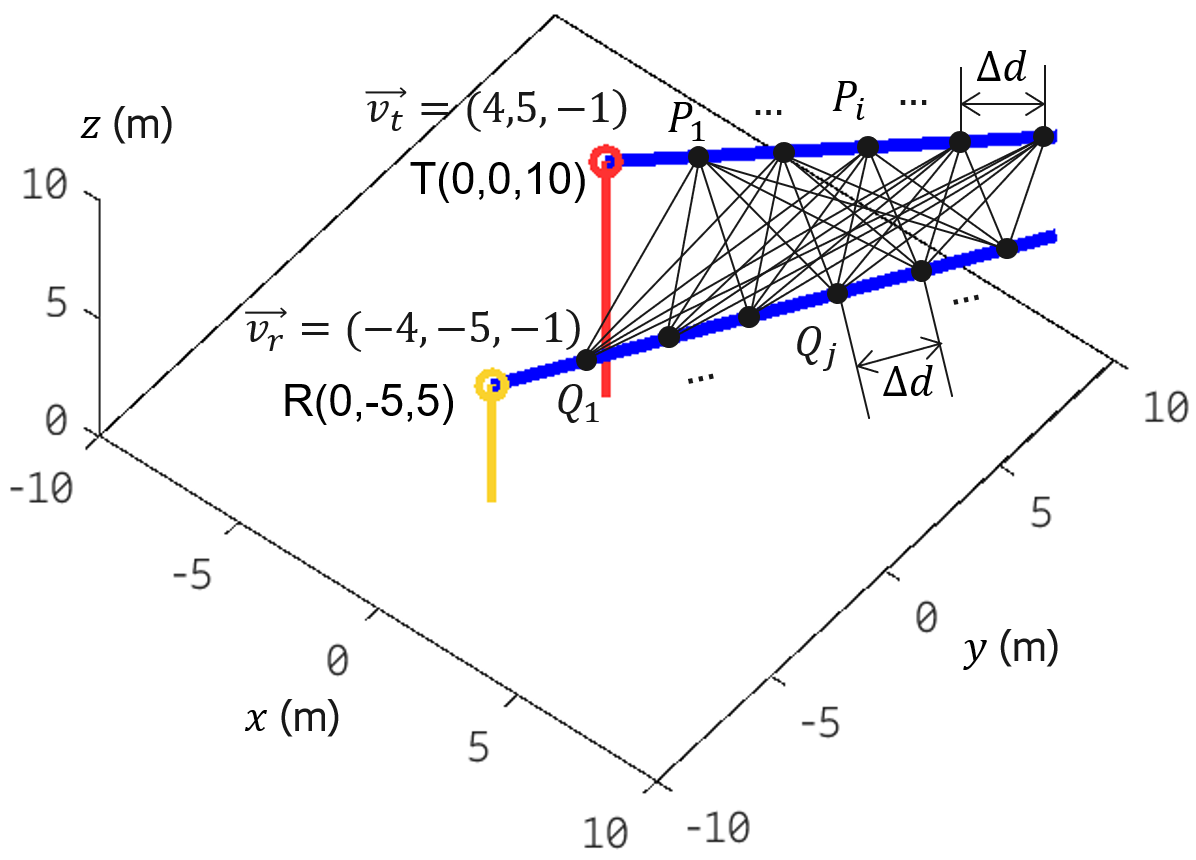}}
	\caption{An illustration of the proposed methods: localize the scatterers and identify their materials by a radio ray transmitted from \mbox{T(0, 0, 10)} in the direction of vector (4, 5, -1) and received at \mbox{R(0, -5, 5)} in the direction of vector (-4, -5, -1).}
	\label{fig_7}
\end{figure}

\begin{figure*}[t]
	\centering
	\subfloat[Intersection of RL data surfaces with plane $z$=25.66]{\label{fig:8a}\includegraphics[width=1\columnwidth]{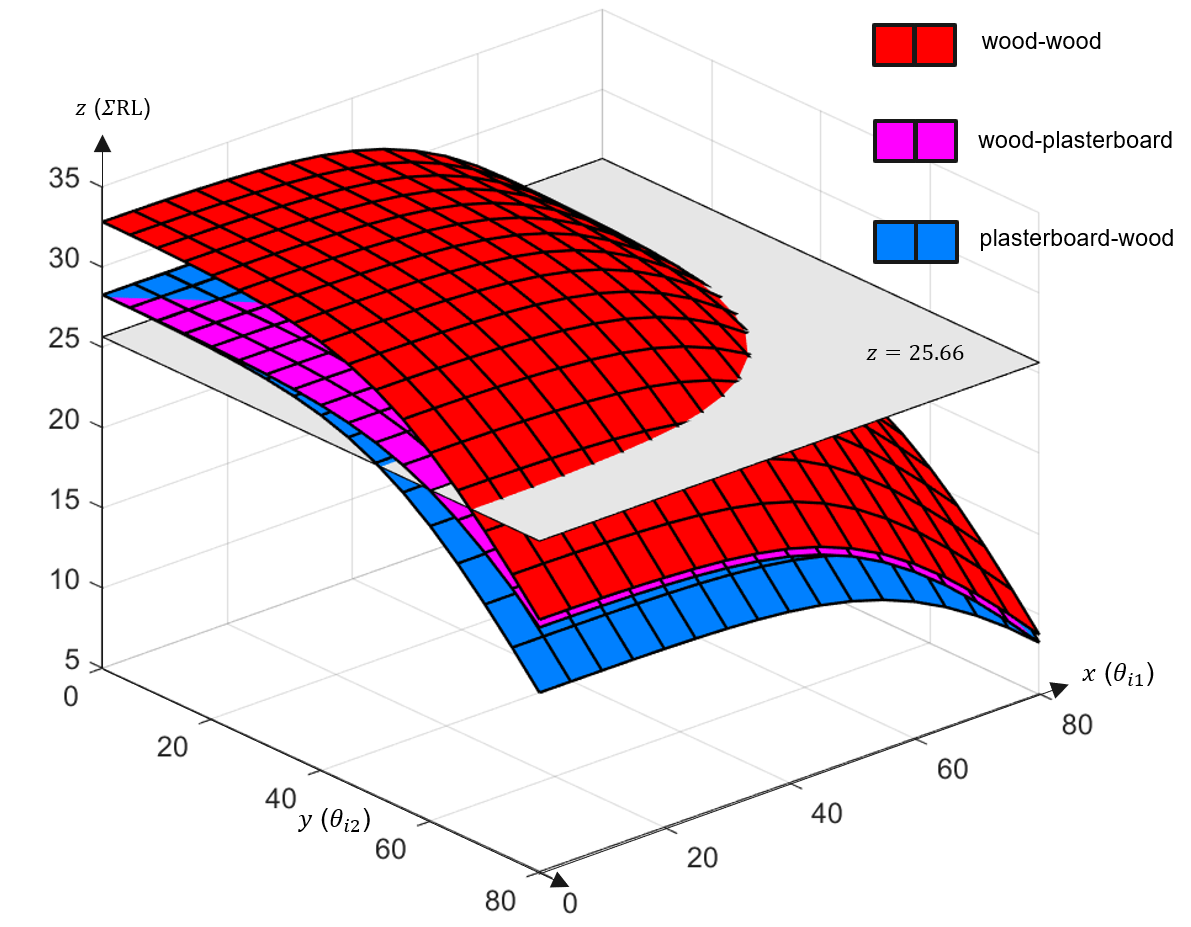}}\quad
	\subfloat[$xy$-traces by letting $z$=25.66 ]{\label{fig:8b}\includegraphics[width=1\columnwidth]{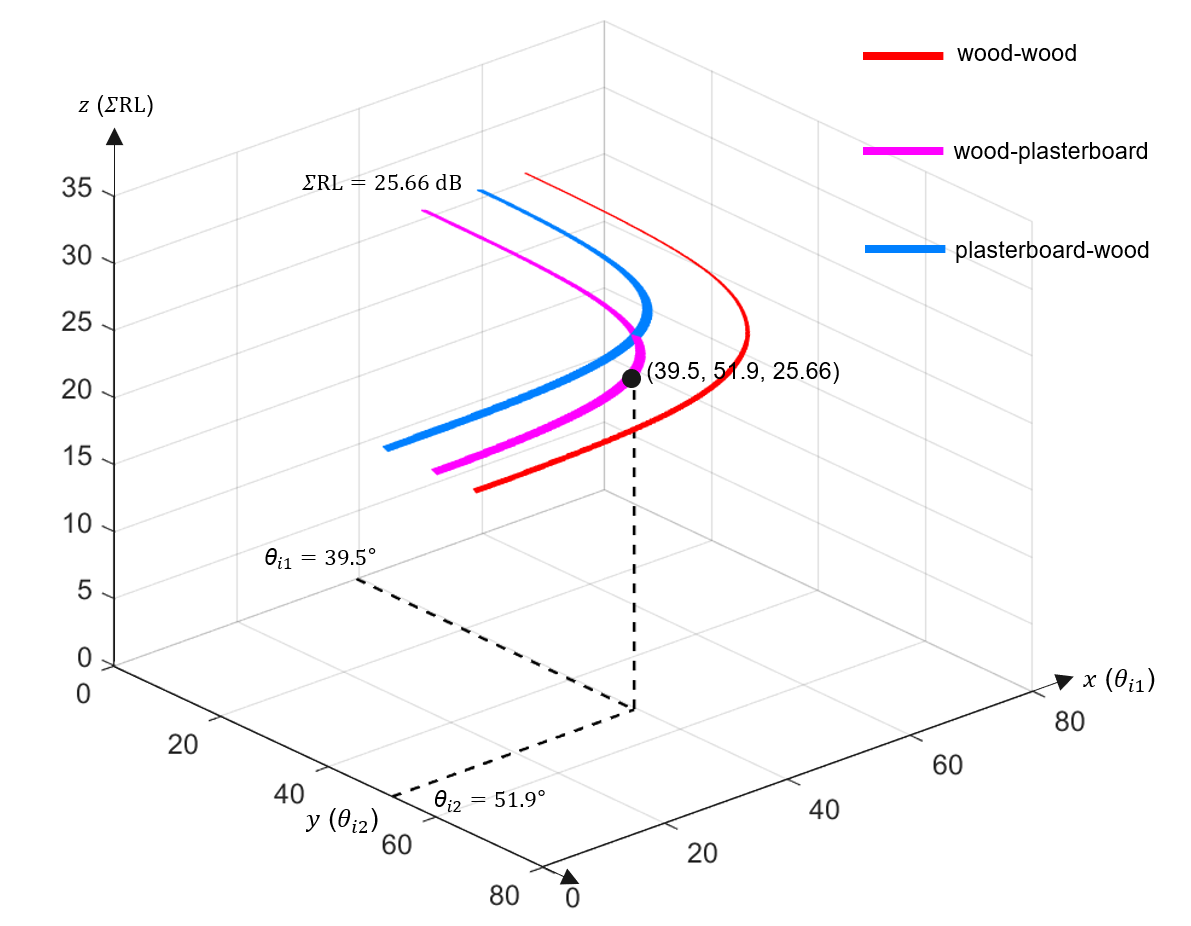}}\\	
	\caption{By intersecting plane $z$=25.66 with RL data surfaces, all possible sequences-of-material and incident angles that induce RL of 25.66~dB can be obtained.}
\end{figure*}

We take the scenario in Fig.~7 as an example to illustrate the proposed method~1. In Step~2, a radio ray transmitted from the TX at \mbox{T(0, 0, 10)} in the direction of vector \mbox{$\overrightarrow{v_t}$=(4, 5, -1)} is reflected by unknown scatterers (i.e., unknown position, unknown number of reflection bounces, and unknown material of scatterers), and the same ray is received by the RX at \mbox{R(0, -5, 5)} in the direction of vector \mbox{$\overrightarrow{v_r}$=(-4, -5, -1)}. We assume that the overall path length is 32.4~m according to TOF measurement, and measured RL or $\Sigma$RL is 22.24~dB at 100~GHz by using the measurement method proposed in Section~III. The goal is to trace the trajectory, localize the RPs, and identify the material of scatterers for this radio ray.

 Vector \mbox{$\overrightarrow{v_t}$=(4, 5, -1)} and vector \mbox{$\overrightarrow{v_r}$=(-4, -5, -1)} are not co-planar, in other words, the transmitter- and receiver-side beam cannot intersect at a point. Therefore, this radio ray is impossible to be a trajectory with single-bounce reflection. In Step~3, we first determine a set of possible trajectories by matching the overall path length $d$ ($d=c\cdot{t}$ where $c$ is the speed of light and $t$ is TOF). The aim is to localize a pair-of-RPs lying on this double-bounce reflection trajectory. To achieve this, we distribute evenly spaced points on the transmitter- and receiver-side-beam ray, respectively. As shown in Fig.~7, a collection of points lying on the transmitter-side-beam ray is defined as set $P=\{P_1, P_2, \ldots, P_i, \ldots\}$, the distance between any adjacent points is $\Delta d$. The selection of $\Delta d$ needs to carefully balance localization accuracy and computational complexity, a shorter $\Delta d$ can improve localization accuracy but leads to a relatively long latency and high computational complexity. The coordinate of any point in set $P$ can be obtained by the coordinate of initial point (0, 0, 10), $\Delta d$, the transmitter-side beam direction $\overrightarrow{v_t}$, and index~$i$. Similarly, we define a set $Q=\{Q_1, Q_2, \ldots, Q_j, \ldots\}$ for receiver-side-beam ray. Then a path-length-matching algorithm is used to identify all possible pairs-of-RPs and all possible trajectories. The algorithm consists of the following procedure: 
 
 For any point $P_i$ in set $P$ and any point $Q_j$ in set $Q$, we calculate the overall double-bounce reflection path length $d(P_i,Q_j)$. $d(P_i,Q_j)$ is the sum of the lengths of the three line segments $\overline{TP_i}$, $\overline{{P_i}{Q_j}}$, and $\overline{Q_jR}$. In case of $d(P_i,Q_j) \neq c\cdot{t}$, we ignore this $(P_i,Q_j)$. Repeat this procedure until $d(P_i,Q_j)$ is longer than or equal to the overall path length of 32.4~m. Finally, four pairs-of-RPs are obtained after the algorithm enumerates all combinations of $P_i$ and $Q_j$:

\begin{itemize}
	\item Pair A: RP1(8, 10, 8), RP2(10, 7.5, 7.5);
	\item Pair B: RP1(8.6, 10.75, 7.85), RP2(6, 2.5, 6.5);
	\item Pair C: RP1(5.8, 7.25, 8.55), RP2(10.9, 8.62, 7.73);
	\item Pair D: RP1(0.8, 1, 9.8), RP2(11.1, 8.88, 7.78).
\end{itemize} 

The four trajectories A-D with double-bounce reflection induced by scatterers locate at pairs-of-RPs A-D are depicted in \mbox{Fig.~2(a)-(d)}, the corresponding incident angles at each RP can be obtained by \eqref{eqn_14} and depicted in \mbox{Fig.~2(a)-(d)} also.

\begin{equation}\label{eqn_14}
	\theta = \frac{1}{2}\arccos(\frac{u\cdot{v}}{\|u\|\|v\|}),
\end{equation}
where $u$ and $v$ are the two Euclidean vectors initialled from RP and along with incident ray and outgoing ray, respectively.

\begin{table*}[h]
	\caption{All possible trajectories, and their RP coordinates, incident angles $\theta_{i1}$ and $\theta_{i2}$, all possible sequences-of-material, RL1 induced by scatterer1, RL2 induce by scatterer2, and $\Sigma$RL (RL1 plus RL2) in case of a radio ray transmitted from \mbox{TX(0, 0, 10)} in the direction of vector (4, 5, -1) and received at \mbox{RX(0, -5, 5)} in the direction of vector (-4, -5, -1) and the overall path length is 32.4~m}
	\begin{center}
		\begin{tabular}{cccccccccc}
			\hline
			Trajectory & RP1 & RP2 & {$\theta_{i1}$} & $\theta_{i2}$ & Scatterer1 & Scatterer2 & RL1 (dB) & RL2 (dB) & $\Sigma $RL \\
			\hline
			\multirow{9}{*}{A} & \multirow{9}{*}{(8, 10, 8)} & \multirow{9}{*}{(10, 7.5, 7.5)} & \multirow{9}{*}{39.5\textdegree} & \multirow{9}{*}{51.9\textdegree} & Wood & Wood & 15.28 & 13.52 & 28.8 \\
			& & & & & Plasterboard & Plasterboard & 11.35 & 10.38 & 21.73 \\
			& & & & & Glass & Glass & 7.29 & 6.96 & 14.25 \\
			& & & & & Wood & Plasterboard & 15.28 & 10.38 & 25.66 \\
			& & & & & Plasterboard & Wood & 11.35 & 13.52 & 24.87 \\
			& & & & & Wood & Glass & 15.28 & 6.96 & 22.24 \\
			& & & & & Glass & Wood & 7.29 & 13.52 & 20.81 \\
			& & & & & Plasterboard & Glass & 11.35 & 6.96 & 18.31 \\
			& & & & & Glass & Plasterboard & 7.29 & 10.38 & 17.67 \\
			\hline
			\multirow{9}{*}{B} & \multirow{9}{*}{(8.6, 10.75, 7.85)} & \multirow{9}{*}{(6, 2.5, 6.5)}& \multirow{9}{*}{13.8\textdegree} & \multirow{9}{*}{79.6\textdegree} & Wood & Wood & 16.34 & 4.34 & 20.68 \\
			& & & & & Plasterboard & Plasterboard & 11.84 & 3.85 & 15.69 \\
			& & & & & Glass & Glass & 7.34 & 3.63 & 10.97 \\
			& & & & & Wood & Plasterboard & 16.34 & 3.85 & 20.19 \\
			& & & & & Plasterboard & Wood & 11.84 & 4.34 & 16.18 \\
			& & & & & Wood & Glass & 16.34 & 3.63 & 19.97 \\
			& & & & & Glass & Wood & 7.34 & 4.34 & 11.68 \\
			& & & & & Plasterboard & Glass & 11.84 & 3.63 & 15.47 \\	
			& & & & & Glass & Plasterboard & 7.34 & 3.85 & 11.19 \\
			\hline
			\multirow{9}{*}{C} & \multirow{9}{*}{(5.8, 7.25, 8.55)} & \multirow{9}{*}{(10.9, 8.62, 7.73)}& \multirow{9}{*}{72.1\textdegree} & \multirow{9}{*}{20.1\textdegree} & Wood & Wood & 7.52 & 16.25 & 23.77 \\
			& & & & & Plasterboard & Plasterboard & 6.37 & 11.81 & 18.18 \\
			& & & & & Glass & Glass & 5.28 & 7.34 & 12.62 \\
			& & & & & Wood & Plasterboard & 7.52 & 11.81 & 19.33 \\
			& & & & & Plasterboard & Wood & 6.37 & 16.25 & 22.62 \\
			& & & & & Wood & Glass & 7.52 & 7.34 & 14.86 \\
			& & & & & Glass & Wood & 5.28 & 16.25 & 21.53 \\
			& & & & & Plasterboard & Glass & 6.37 & 7.34 & 13.71 \\
			& & & & & Glass & Plasterboard & 5.28 & 11.81 & 17.09 \\
			\hline
			\multirow{9}{*}{D} & \multirow{9}{*}{(0.8, 1, 9.8)} & \multirow{9}{*}{(11.1, 8.88, 7.78)}& \multirow{9}{*}{83.1\textdegree} & \multirow{9}{*}{11.3\textdegree} & Wood & Wood & 3.08 & 16.37 & 19.45 \\
			& & & & & Plasterbaord & Plasterboard & 2.78 & 11.85 & 14.63 \\
			& & & & & Glass & Glass & 2.76 & 7.34 & 10.1 \\
			& & & & & Wood & Plasterboard & 3.08 & 11.85 & 14.93 \\
			& & & & & Plasterboard & Wood & 2.78 & 16.37 & 19.15 \\
			& & & & & Wood & Glass & 3.08 & 7.34 & 10.42 \\
			& & & & & Glass & Wood & 2.76 & 16.37 & 19.13 \\
			& & & & & Plasterboard & Glass & 2.78 & 7.34 & 10.12 \\
			& & & & & Glass & Plasterboard & 2.76 & 11.85 & 14.61 \\
			\hline
		\end{tabular}
	\end{center}
	\label{tab_3}
\end{table*}

In Step~4, the true pair-of-RPs is identified by comparing the measured $\Sigma$RL with the data in the RL database created in Step~1. RL induced by the first scatterer (RL1) at incident angle $\theta_{i1}$, RL induced by the second scatterer (RL2) at incident angle $\theta_{i2}$ of trajectories A-D are summarized in Table~III. From the $\Sigma$RL data in Table~III, it is evident that there are significant differences between the most of RLs of trajectories A-D with different sequences-of-material. As mentioned earlier, a RL value of 22.24~dB has already been measured at 100~GHz in Step~2. By looking up Table~III, a row with $\Sigma$RL=22.24~dB is found. From the information contained in this row, we can conclude that trajectory A is the true trajectory, RP1 locates at \mbox{(8, 10, 8)} and the first scatterer is made of wood, RP2 locates at \mbox{(10, 7.5, 7.5)} and the second scatterer is made of glass. The information of the two RPs can be used to represent two points in 3D space with \mbox{4-dimensional} information including Cartesian coordinate~($x, y, z$) and material information. By collecting enough points in the environment and converting the point cloud to 3D surface, a 3D digital map with material information can be generated.

An alternative method (referred to as method~2 below) for scatterer localization and material identification can be developed by executing Step~3 and 4 of method~1 in reverse order. Similar to method~1, method~2 carries out a four-step procedure:

\begin{itemize}
	\item Step 1 and 2: essentially the same as method~1.
	\item Step 3: find all sequences-of-material and incident angles from the RL database such that the RL induced by scatterers with these sequences-of-material and incident angles is close to measured RL.
	\item Step 4: estimate the positions of RPs of every possible trajectory by matching the overall path length.
\end{itemize} 

\begin{table*}[h]
	\caption{All possible trajectories, and their RP coordinates, incident angles $\theta_{i1}$ and $\theta_{i2}$, all possible sequences-of-material, and overall path length in case of a radio ray transmitted from \mbox{TX(0, 0, 10)} in the direction of vector (4, 5, -1) and received at \mbox{RX(0, -5, 5)} in the direction of vector (-4, -5, -1) and the scatterers induce RL of 25.66~dB}
	\begin{center}
		\begin{tabular}{cccccccc}
			\hline
			Trajectory & RP1 & RP2 & {$\theta_{i1}$} & $\theta_{i2}$ & Scatterer1 & Scatterer2 & path length (m) \\
			\hline
			1 & (11, 13.75, 7.25) & (16, 15, 9) & 67\textdegree & 18.7\textdegree & wood & wood & 49.2 \\
			2 & (13, 16.25, 6.75) & (17, 16.25, 9.25) & 58.1\textdegree & 26.4\textdegree & plasterboard & wood & 53.3 \\
			3 & (7, 8.75, 8.25) & (10, 7.5, 7.5) & 53.7\textdegree & 38.3\textdegree & plasterboard & wood & 30.9 \\
			4 & (14, 17.5, 6.5) & (17, 16.25, 9.25) & 48.1\textdegree & 36\textdegree & wood & plasterboard & 54.5 \\
			5 & (8, 10, 8) & (10, 7.5, 7.5) & 39.5\textdegree & 51.9\textdegree & wood & plasterboard & 32.4 \\
			6 & (12, 15, 7) & (14, 12.5, 8.5) & 37.4\textdegree & 48.8\textdegree & plasterboard & wood & 45.7 \\
			7 & (14, 17.5, 6.5) & (15, 13.75, 8.75) & 27.1\textdegree & 57.7\textdegree & wood & plasterboard & 51.5 \\
			8 & (13, 16.25, 6.75) & (13, 11.25, 8.25) & 19.2\textdegree & 67\textdegree & wood & wood & 47.3 \\
			\hline
		\end{tabular}
	\end{center}
	\label{tab_4}
\end{table*}

\begin{figure}[t]
	\centerline{\includegraphics[width=\linewidth, height=10cm, keepaspectratio]{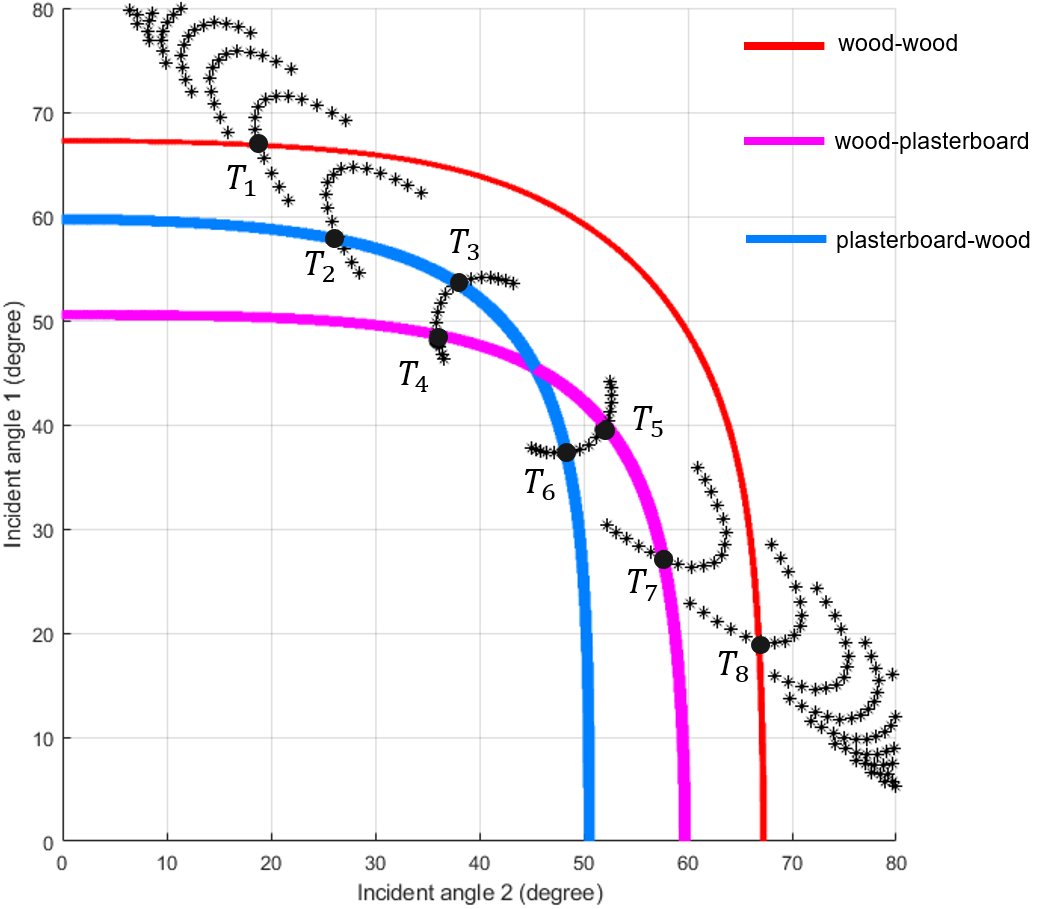}}
	\caption{An illustration of how the true trajectory is identified by Step~4 of method~2.}
	\label{fig_9}
\end{figure}

Step~1 and 2 of method~2 just do the same as the steps of method~1, namely to create a RL database and then measure the RL of a trajectory in the environment to be detected. In Step~3, RL matching is performed, in other words, each RL in the RL database is compared with the measured RL and the rows with same RL as measured RL are identified. The purpose of RL matching is to find all possible sequences-of-material and incident angles that induce same RL as measured RL. To help visualize the possible sequences-of-material and their incident angles, the trace of the RL data surfaces in Fig.~4(b) is plotted. The trace of a surface is the cross section of the surface with a plane parallel to one of the coordinate planes. For example, we assume that the measured RL is 25.66~dB in the scenario depicted in Fig.~7. As shown in Fig.~8(a), by intersecting plane $z$ = 25.66 with the double-bounce RL data surfaces, we find three traces. Removing the RL data surfaces helps us visualize the three traces and they are illustrated in Fig.~8(b). These three traces consist of all sequences-of-material and incident angles information about trajectories that can induce RL of 25.66~dB. For example, in Fig.~8(b), point (39.5, 51.9, 25.66) lying on the trace curve colored with magenta indicates that a double-bounce-reflection trajectory may contain the following propagation properties: (1) the first reflection is induced by a scatterer made of wood at incident angle $\theta_{i1}$=39.5°; (2) the second reflection is induced by a scatterer made of plasterboard at incident angle $\theta_{i2}$=51.9°; (3) the $\Sigma$RL of this trajectory is 25.66~dB. By Step~3, we can find all possible sequences-of-material and incident angles of double-bounce reflection with RL of 25.66~dB.

In Step~4, we first calculate the incident angles of all possible trajectories using AOD, AOA, and coordinates of TX and RX. Similar to Step~3 of method~1, as shown in Fig.~7, for any point $P_i$ in set $P$ and any point $Q_j$ in set $Q$, joining T(0, 0, 10), $P_i$, $Q_j$, and R(0, -5, 5) in sequence can trace a trajectory. The first incident angle $\theta_{i1}$ ($\theta_{i1}=\widehat{TP_i{Q_j}}/2$) and the second incident angle $\theta_{i2}$ ($\theta_{i2}=\widehat{P_i{Q_j}R}/2$) of this trajectory can be obtained by using \eqref{eqn_14}. In Fig.~9, we show the incident angles $\theta_{i1}$ and $\theta_{i2}$ of every possible trajectories by enumerates all combinations of $P_i$ and $Q_j$ until the trajectory length is longer than 32.4~m. The incident angles of each trajectory is marked by a star (*), the $y$- and $x$-coordinate of a star marker represents the first incident angle $\theta_{i1}$ and the second incident angle $\theta_{i2}$ of a trajectory, respectively. If we combine the three traces in Fig.~8(b) into Fig.~9, the traces intersect the black star markers at several points $T_1, T_2, \ldots, T_8$. These points represent eight trajectories 1-8 that satisfy the following conditions: (1) the RPs induce $\Sigma$RL of 25.66~dB; (2) the RPs lie on the transmitter- and the receiver-side beam ray, respectively. Table~IV summarizes the coordinates of RPs, incident angles, and sequence-of-material of trajectories 1-8. The overall path length of trajectories 1-8 are also calculated by coordinates of RPs. Finally, by comparing the path length of trajectories 1-8 with the measured path length of 32.4~m, it can be concluded that trajectory~5 is the true trajectory.

\subsection{Performance evaluation}
The proposed methods localize scatterers by RL which is obtained by measured RSS. Therefore, their performance is highly affected by accuracy of RSS measurement (RSS uncertainty) \cite{b5}. RSS uncertainty is influenced by some deterministic and stochastic factors \cite{b12}. The deterministic factors can be predicted and estimated, such as antenna patterns and measurement instruments’ resolution. For example, by using high performance equipment (e.g., massive multiple-input multiple-output (MIMO) antennas) and measurement instruments with highly accurate power resolution, the RSS measurement error can be minimized \cite{b13}. Unlike deterministic factors, stochastic factors are time-varying and hard to be estimated before measurement \cite{b12}, RSS uncertainty induced by stochastic factors may result from non-ideal equipment, hardware impairment, Gaussian random noise, and interference from neighboring antennas in a JCAS network. To eliminate RSS uncertainty caused by stochastic factors, we can perform multiple RSS measurements, then calculate the mean RSS and subsequently estimate the RL to be used in the proposed methods \cite{b12}. With the help of multiple measurements, a relatively long measurement period can achieve better SNR and lower standard error of RSS uncertainty than RSS uncertainty by single measurement. Simulation in \cite{b12} shows that the standard error of Gaussian random noise with 0~dB mean and 0.5~dB standard deviation decreased to 0.05~dB when 50 measurements have been carried out. The RSS-Measurement-Muting (RMM) can be used to mitigate interference from the environment for RSS measurement also. RMM delays or avoids the transmission of any messages of neighboring nodes for a time period when a RSS measurement is performed by a JCAS node.

It is expected that higher accuracy of RSS measurement will be achieved in 6G. In particular, RSS-based JCAS applications have a definite need for accurate measurement to sense and identify objects. Higher accuracy of measurement can improve the performance of the proposed methods. For example, we assume that the measured RL is 12.7~dB in the scenario illustrated in Fig.~7, and the RSS uncertainty of the JCAS scatterer localization system is 1.5~dB. In other words, the true RL of this trajectory ranges from 11.2 to 14.2~dB (12.7$\pm$1.5~dB). By looking up the RL data in Table~III, three RLs (11.68~dB, 12.62~dB, and 13.71~dB) are within the range. Therefore, we can conclude that the coordinates and materials of the two scatterers could be one of the below cases:

\begin{itemize}
	\item glass at (8.6, 10.75, 7.85), wood at (6, 2.5, 6.5);
	\item glass at (5.8, 7.25, 8.55), glass at (10.9, 8.62, 7.73);
	\item plasterboard at (5.8, 7.25, 8.55), glass at (10.9, 8.62, 7.73).
\end{itemize} 

By taking advantage of the 6G technologies, we can assume that the performance of measurement can be improved, e.g., RSS uncertainty can be reduced to 1~dB. Now the true RL in previous example ranges from 11.7 to 13.7~dB (12.7$\pm$1~dB). We can immediately identify that the first scatterer is made of glass and locates at \mbox{(5.8, 7.25, 8.55)}, the second scatterer is made of glass also and locates at \mbox{(10.9, 8.62, 7.73)}. Because only one RL of 12.62~dB in Table~III is within the range of 12.7$\pm$1~dB.

\section{Conclusions}
Reusing radio signals of cellular network to localize scatterers will potentially replace some functionalities of radar/LiDAR in 6G, for example, 3D mapping. However, there has been little quantitative research on this topic so far. The process of scatterer localization can be simplified by making single-bounce-assumption as workaround, but it cannot meet the extreme reliability requirements of some critical use cases, e.g., AD. In this article, we extend the capability of the method proposed in \cite{b8} from material identification only to both material identification and scatterer localization simultaneously. The extension is based on findings from RL simulations and theoretical analysis: (1) most of the $\Sigma$RLs induced by different sequences-of-material are significantly different; (2) RL is dependent on material of scatterer, radio frequency, and incident angle only; (3) $\Sigma$RLs induced by multiple scatterers is independent of scatterers' sequence-of-material. By extracting material and incident angle information from RL, the RAT-based scatterer localization and material identification systems can be developed. The proposed methods will be able to support passive localization in a cost effective manner by eliminating the need for dedicated hardware (e.g., sensor) and external assistance (e.g., 3D digital map database) by using only cellular infrastructure.

\section*{Acknowledgment}
This work has been partly funded by the European Commission through the H2020 project Hexa-X (Grant Agreement no. 101015956).

\vspace{12pt}
\color{red}
\end{document}